\documentclass[format=acmsmall, review=false, screen=true]{acmart}
\usepackage{textcomp}
\usepackage{booktabs}
\usepackage{enumitem}
\usepackage{multirow}
\usepackage{subfig}
\usepackage{listings}
\usepackage{color}

\definecolor{dkgreen}{rgb}{0,0.5,0}
\definecolor{dkred}{rgb}{0.5,0,0}
\definecolor{gray}{rgb}{0.5,0.5,0.5}
\begin{document}
\title{psc2code: Denoising Code Extraction from Programming Screencasts}

\author{Lingfeng Bao}
\affiliation{%
\institution{College of Computer Science and Technology, Zhejiang University}
\streetaddress{38 Zheda Rd}
\city{Hangzhou}
\state{Zhejiang}
\postcode{310000}
\country{China}}
\affiliation{%
\institution{Ningbo Research Institute, Zhejiang University}
\city{Ningbo}
\state{Zhejiang}
\country{China}}
\affiliation{%
\institution{PengCheng Laboratory}
\city{Shenzhen}
\country{China}}
\email{lingfengbao@zju.edu.cn}
\author{Zhenchang Xing}
\affiliation{%
  \institution{Australian National University}
  \streetaddress{Acton ACT 2601}
  \city{Canberra}
  \country{Australia}}
\email{zhenchang.Xing@anu.edu.au}
\author{Xin Xia}
\authornote{Corresponding Authors: Xin Xia}
\affiliation{%
 \institution{Monash University}
 \streetaddress{Wellington Rd}
 \city{Melbourne}
 \state{Victoria}
 \country{Australia}}
\email{Xin.Xia@monash.edu}
\author{David Lo}
\affiliation{%
  \institution{Singapore Management University}
  \streetaddress{81 Victoria St}
  \postcode{188065}
  \country{Singapore}}
\email{davidlo@smu.edu.sg}
\author{Minghui Wu}
\affiliation{%
  \institution{Zhejiang University City College}
  \streetaddress{51 Huzhou Rd}
  \city{Hangzhou}
  \state{Zhejiang}
  \postcode{310000}
  \country{China}}
\email{mhwu@zucc.edu.cn}
\author{Xiaohu Yang}
\affiliation{%
  \institution{Zhejiang University}
  \streetaddress{38 Zheda Rd}
  \city{Hangzhou}
  \state{Zhejiang}
  \postcode{310000}
  \country{China}}
\email{yangxh@zju.edu.cn}

\renewcommand\shortauthors{L. Bao et al.}

\begin{abstract}
  Programming screencasts have become a pervasive resource on the Internet, which help developers learn new programming technologies or skills.
  The source code in programming screencasts is an important and valuable information for developers.
  But the streaming nature of programming screencasts (i.e., a sequence of screen-captured images) limits the ways that developers can interact with the source code in the screencasts.
  Many studies use the Optical Character Recognition (OCR) technique to convert screen images (also referred to as video frames) into textual content, which can then be indexed and searched easily.
  However, noisy screen images significantly affect the quality of source code extracted by OCR, for example, no-code frames (e.g., PowerPoint slides, web pages of API specification), non-code regions (e.g., Package Explorer view, Console view), and noisy code regions with code in completion suggestion popups.
  Furthermore, due to the code characteristics (e.g., long compound identifiers like ItemListener), even professional OCR tools cannot extract source code without errors from screen images.
  The noisy OCRed source code will negatively affect the downstream applications, such as the effective search and navigation of the source code content in programming screencasts.

  In this paper, we propose an approach named \emph{psc2code} to denoise the process of extracting source code from programming screencasts.
  First, \emph{psc2code} leverages the Convolutional Neural Network (CNN) based image classification to remove non-code and noisy-code frames.
  Then, \emph{psc2code} performs edge detection and clustering-based image segmentation to detect sub-windows in a code frame, and based on the detected sub-windows, it identifies and crops the screen region that is most likely to be a code editor.
  Finally, \emph{psc2code} calls the API of a professional OCR tool to extract source code from the cropped code regions and leverages the OCRed cross-frame information in the programming screencast and the statistical language model of a large corpus of source code to correct errors in the OCRed source code.

  We conduct an experiment on 1,142 programming screencasts from YouTube.
  We find that our CNN-based image classification technique can effectively remove the non-code and noisy-code frames, which achieves a F1-score of 0.95 on the valid code frames.
  We also find that \emph{psc2code} can significantly improve the quality of the OCRed source code  by truly correcting about half of incorrectly OCRed words.
  Based on the source code denoised by \emph{psc2code}, we implement two applications:
  1) a programming screencast search engine; 2) an interaction-enhanced programming screencast watching tool.
  Based on the source code extracted from the 1,142 collected programming screencasts, our experiments show that our programming screencast search engine achieves the precision@5, 10, and 20 of 0.93, 0.81, and 0.63, respectively.
  We also conduct a user study of our interaction-enhanced programming screencast watching tool with 10 participants.
  This user study shows that our interaction-enhanced watching tool can help participants learn the knowledge in the programming video more efficiently and effectively.
\end{abstract}

\setcopyright{acmcopyright}
\acmJournal{TOSEM}
\acmYear{2019} \acmVolume{1} \acmNumber{1} \acmArticle{1} \acmMonth{1} \acmPrice{15.00}



\maketitle

\section{Introduction}\label{sec:intro}
Programming screencasts, such as programming video tutorials on YouTube, can be recorded by screen-capturing tools like Snagit~\cite{Snagit}.
They provide an effective way to introduce programming technologies and skills, and offer a live and interactive learning experience.
In a programming screencast, a developer can teach programming by developing code on-the-fly or showing the pre-written code step by step.
A key advantage of programming screencasts is the viewing of a developer's coding in action, for example, how changes are made to the source code step-by-step and how errors occur and are being fixed~\cite{macleod2015code}.

There are a huge number of programming screencasts on the Internet.
For example, YouTube, the most popular video-sharing website, hosts millions of programming video tutorials.
The Massive Open Online Course (MOOC) websites (e.g. {Coursera\footnote{\url{https://www.coursera.org}}, edX\footnote{\url{https://www.edx.org}}) and the live streaming websites (e.g. Twitch\footnote{\url{https://www.twitch.tv/}}) also provide many resources of programming screencasts.
However, the streaming nature of programming screencasts, i.e., a stream of screen-captured images, limits the ways that developers can interact with the content in the videos.
As a result, it can be difficult to search and navigate programming screencasts.

To enhance the developer's interaction with programming screencasts, an intuitive way is to convert video content into text (e.g., source code) by the Optical Character Recognition (OCR) technique.
As textual content can be easily indexed and searched, the OCRed textual content makes it possible to find the programming screencasts with specific code elements in a search query.
Furthermore, video watchers can quickly navigate to the exact point in the screencast where some APIs are used.
Last but not the least, the OCRed code can be directly copied and pasted to the developer's own program.

However, extracting source code accurately from programming screencasts has to deal with three ``noisy'' challenges (see Section~\ref{sec:motivation} for examples).
First, developers in programming screencasts not only develop code in IDEs (e.g., Eclipse, Intellij IDEA) but also use some other software applications, for example, to introduce some concepts in power point slides, or to visit some API specifications in web browsers.
Such non-code content does not need to be extracted if one is only interested in the code being developed.
Second, in addition to code editor, modern IDEs include many other parts (e.g., tool bar, package explorer, console, outline, etc.).
Furthermore, the code editor may contains popup menu, code completion suggestion window, etc.
The mix of source code in code editor and the content of other parts of the IDE often result in poor OCR results.
Third, even for a clear code editor region, the OCR techniques cannot produce 100\% accurate text due to the low resolution of screen images in programming screencasts and the special characteristics of GUI images (e.g., code highlights, the overlapping of UI elements).

Several approaches have been proposed to extract source code from programming screencasts~\cite{bao2015reverse, ponzanelli2016too, yadid2016extracting, khandwala2018codemotion}.
A notable work is CodeTube~\cite{ponzanelli2016too, ponzanelli2017automatic}, a programming video search engine based on the source code extracted by the OCR technique.
One important step in CodeTube is to extracts source code from programming screencasts. It recognizes the code region in the frames using the computer vision techniques including shape detection and frame segmentation, followed by extracting code constructs from the OCRed text using an island parser.

However, CodeTube does not explicitly address the aforementioned three ``noisy'' challenges.
First, it does not distinguish code frames from non-code frames before the OCR.
Instead, it OCRs all the frames and check the OCRed results to determine whether a frame contains the code.
This leads to unnecessary OCR for non-code frames.
Second, CodeTube does not remove noisy code frames, for example, the frames with code completion suggestion popups.
Not only is the quality of the OCRed text for this type of noisy frames low, but also the OCRed text highly likely contains code elements that appear only in popups but not in the actual program.
Third, CodeTube simply ignores the OCR errors in the OCRed code using a code island parser, and does not attempt to fix the OCR errors in the output code.

In this work, we propose \emph{psc2code}, a systematic approach and the corresponding tool that explicitly addresses the three ``noisy'' challenges in the process of extracting source code from programming screencasts. 
First, \emph{psc2code} leverages the Convolutional Neural Network (CNN) based image classification to remove frames that have no code and noisy code (e.g., code is partially blocked by menus, popup windows, completion suggestion popups, etc.) before OCRing code in the frames.
Second, \emph{psc2code} attempts to distinguish code regions from non-code regions in a frame.
It first detects Canny edges~\cite{canny1986computational} in a code frame as candidate boundary lines of sub-windows.
As the detected boundary lines tend to be very noisy, \emph{psc2code} clusters close-by boundary lines and then clusters frames with the same window layout based on the clustered boundary lines.
Next, it uses the boundary lines shared by the majority of the frames in the same frame cluster to detect sub-windows, and subsequently identify the code regions among the detected sub-windows.
Third, \emph{psc2code} uses the Google Vision API~\cite{googlevision} for text detection to OCR a given code region image into text.
It fixes the errors in the OCRed source code, based on the cross-frame information in the programming screencast and the statistical language model of a large corpus of source code.

To evaluate our proposed approach, we collect 23 playlists with 1142 Java programming videos from YouTube.
We randomly sample 4820 frames from 46 videos (two videos per playlist) and find that our CNN-based model achieves 0.95 and 0.92 F1-score on classifying code frames and non-code/noisy-code frames, respectively.
The experiment results on these sampled frames also show that \emph{psc2code} correct about half of incorrectly-OCRed words (46\%), and thus it can significantly improve the quality of the OCRed source code.

We also implement two downstream applications based on the source code extracted by \emph{psc2code}:
\begin{enumerate}[leftmargin=*]
    \item We build a \emph{programming video search engine} based on the source code of the 1142 collected YouTube programming videos. We design 20 queries that consist of commonly-used Java classes or APIs to evaluate the constructed video search engine. The experiment shows that the average precision@5, 10, and 20 are 0.93, 0.81, and 0.63, respectively, while the average precision@5, 10, and 20 achieved by the search engine built on CodeTube~\cite{ponzanelli2016too} are 0.53, 0.50, and 0.46, respectively.

    \item We implement an interaction-enhanced tool for watching programming screencasts. The interaction features include navigating the video by code content, viewing file content, and action timeline.
    We conduct a user study with 10 participants and find that our interaction-enhanced video player can help participants learn the knowledge in the video tutorial more efficiently and effectively, compared with participants using a regular video player.
\end{enumerate}

\vspace{0.1cm} \noindent \textbf{Paper contributions}:
\begin{itemize}
  \item We identify three ``noisy'' challenges in the process of extracting source code from programming screencasts.
  \item We propose and implement a systematic denoising approach to address these three ``noisy'' challenges.
  \item We conduct large-scale experiments to evaluate the effectiveness of our denoising approach and its usefulness in two downstream applications.
\end{itemize}

\vspace{0.1cm} \noindent \textbf{Paper Structure}: The remainder of the paper is structured
as follows.
Section~\ref{sec:motivation} describes the motivation examples of our work.
Section~\ref{sec:approach} presents the design and implementation of \emph{psc2code}.
Section~\ref{sec:experiment} describes the experiment setup and results of \emph{psc2code}.
Section~\ref{sec:app} demonstrates the usefulness of \emph{psc2code} in the two downstream applications based on the source code extracted by \emph{psc2code}.
Section~\ref{sec:diss} discusses the threats to validity in this study.
Section~\ref{sec:related} reviews the related work.
Section~\ref{sec:conclusion} concludes the paper and discusses our future plan.

\section{Motivation}\label{sec:motivation}


We identify three ``noisy'' challenges that affect the process of extracting source code from programming screencasts and the quality of the extracted source code.
In this section, we illustrate these three challenges with examples.

\begin{figure}[t]
    \centering
    \includegraphics[width=0.9\textwidth]{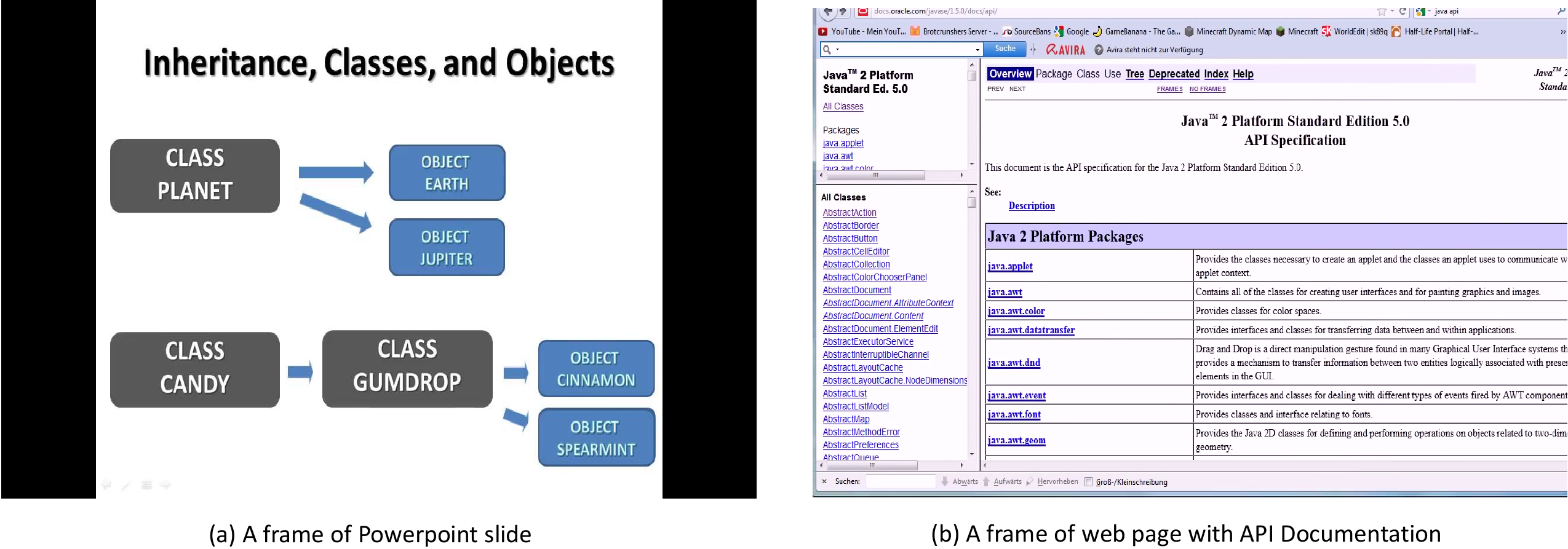}
    \caption{Examples of non-code frames}\label{fig:nocode}
\end{figure}

\subsection{Non-Code Frames}
Non-code frames refer to screen images of software applications other than software development tools or screen images of development tools containing no source code.
Fig.~\ref{fig:nocode} shows some typical examples of non-code frames that we commonly see in YouTube programming videos, including a frame of PowerPoint slide and a frame of web page with API Documentation.
Many non-code frames, such as PowerPoint slides, do not contain source code.
Some non-code frames may contain some code elements and code fragments, for example, API declarations and sample code in the Javadoc pages, or file, class and method names in Package Explorer and Outline views of IDEs.
In this study, we focus on the source code viewed or written by developers in software development tools. Thus, these non-code frames are excluded.

Existing approaches~\cite{ponzanelli2016too,yadid2016extracting,khandwala2018codemotion} blindly OCR both non-code frames and code frames, and then rely on post-processing of the OCRed content to distinguish non-code frames from code frames.
This leads to two issues.
First, the OCR of non-code frames is completely unnecessary and wastes much computing resource and processing time.
Second, the post-processing may retain the code elements in non-code frames which are never used in the programming screencast.
For example, none of the API information in the Javadoc page in Fig.~\ref{fig:nocode}(b) is relevant to the programming screencast discussing the usage of the Math APIs\footnote{\url{https://www.youtube.com/watch?v=GgYXEFhPhRE}}.
Retaining such irrelevant code elements will subsequently result in inaccurate search and navigation of the source code in the programming screencast.
To avoid these two issues, our approach distinguish non-code frames from code frames before the OCR (see Section~\ref{sec:step2}).

\begin{figure}[t]
    \centering
    \includegraphics[width=0.9\textwidth]{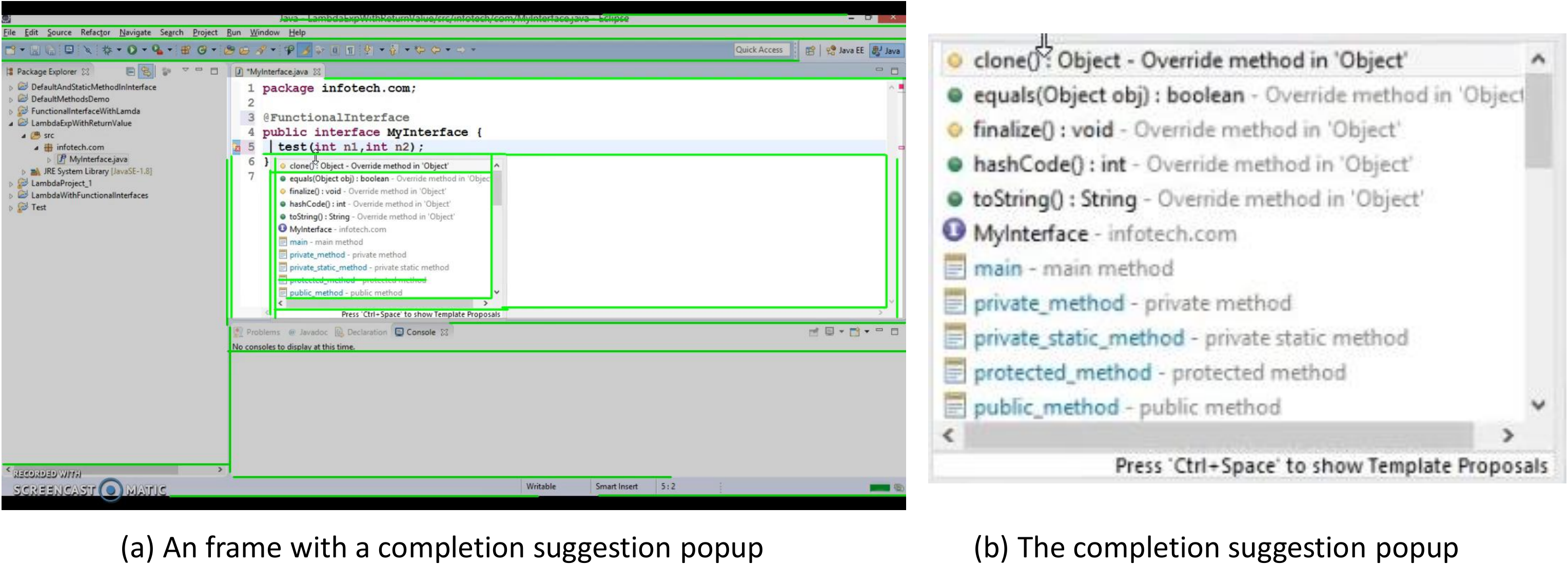}
    \caption{An frame with a completion suggestion popup}\label{fig:invalidated}
\end{figure}

\subsection{Non-Code Regions and Noisy Code Regions}
Modern development tools usually consist of many non-code sub-windows (e.g., Package Explorer, Outline, Console) in addition to the code editor.
As such, a code frame usually contains many non-code regions in addition to the code editor region.
Such UI images consist of multiple regions with independent contents. They are very different from the images that the OCR techniques commonly deal with.
As such, directly applying the OCR techniques to such UI images often results in poor OCR results\footnote{Please refer to the OCR results generated by Google Vision API: \url{https://goo.gl/69a8Vo}}.
Existing approaches~\cite{ponzanelli2016too,yadid2016extracting,khandwala2018codemotion} for extracting source code from programming screencasts leverage the computer vision technique (e.g., edge detection)
to identify the region of interest (ROI) in a frame, which is likely the code editor sub-window in an IDE, and then OCR only the code editor region.

However, as indicated by the green lines in Fig.~\ref{fig:invalidated}(a) (see also the examples in Fig.~\ref{tbl:example}), edge detection on UI images tends to produce very noisy results, due to the presence of multiple sub-windows, scroll bars, code highlights, etc. in the UI images.
The detected noisy horizontal and vertical lines will negatively affect the accurate detection of sub-window boundaries, which in turn will result in inaccurate segmentation of code editor region for OCR.
Existing approaches do not explicitly address this issue.
In contrast, our approach performs edge clustering and frame layout clustering to reduce the detected noisy horizontal and vertical lines and improve the accuracy of sub-window segmentation (see Section~\ref{sec:step3}).

A related issue is noisy code region in which code editor contains some popup window.
For example, when recording a programming screencast, the developer usually writes code on the fly in the IDE, during which many code completion suggestion popups may appear.
As illustrated in the example in Fig.~\ref{fig:invalidated}(b), such popup windows cause three issues.
First, the presence of popup windows complicates the identification of code editor region, as they also contain code elements.
Second, the popup windows may block the actual code in the code editor, and the code elements in popup windows are of different visual presentation and alignment from the code in the code editor.
This may result in poor OCR results. 
Third, popup windows often contains code elements that are never used in the programming screencasts (e.g., the API {\tt hashCode}, {\tt toString} in the popup window in Fig.~\ref{fig:invalidated}).

Existing approaches simply OCR code regions with popup windows.
Not only is the quality of the OCR results low, but it is also difficult to exclude code elements in popup windows from the OCRed text by examining whether the OCR text is code-like or not (e.g., using the island parser in~\cite{ponzanelli2016too}).
The presence of such irrelevant code elements will negatively affects the search and navigation of the source code in programming screencasts.
Addtionally, excluding these noisy-code frames wouldn't lose much important information because we can usually find other frames with similar content in the videos.
In our approach, we consider frames containing code editor with popup window as noisy code frames and exclude noisy code frames using an image classification technique before the OCR (see Section~\ref{sec:step2}).


\begin{figure}[t]
    \centering
    \includegraphics[width=0.6\textwidth]{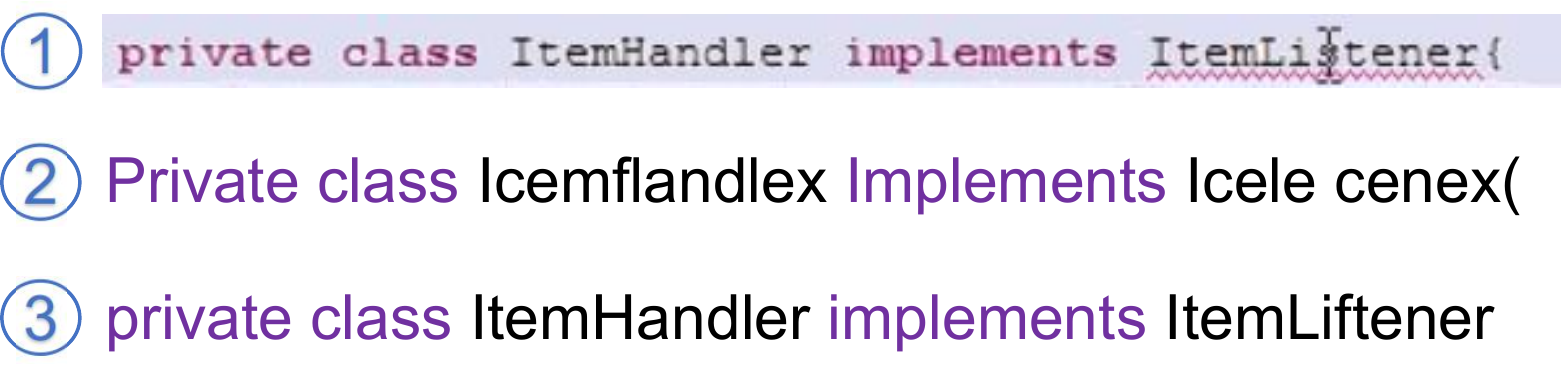}
    \caption{A cropped image of a line of code (1) and its OCR text by Tesseract (2) and Google Vision API (3).}\label{fig:lineocr}
\end{figure}

\subsection{OCR Errors}
Even when the code editor region can be segmented accurately, the code extracted by an OCR technique still typically contain OCR errors, due to three reasons.
First, the OCR techniques generally require the input images with 300 DPI (Dots Per Inch), but the frames in the programming screencasts usually have much lower DPI.
Second, the code highlighting changes the foreground and background color of the highlighted code. This may result in low contrast between the highlighted code and the background, which has an impact of the quality of OCR results.
Third, the overlapping of UI components (e.g., cursor) and the code beneath it may result in OCR errors.

Fig.~\ref{fig:lineocr} shows an example for the second and third reasons.
We use the Tesseract OCR engine~\cite{tesseract} and Google Vision API for text detection~\cite{googlevision} to extract code from a given code region image.
The Tesseract OCR engine is an open source tool developed by Google, and the Google Vision API is a professional computation vision service provided by Google that supports image labeling, face, logo and landmark detection, and OCR.
As seen in the OCR results, both Tesseract and Google Vision API fail to correctly OCR ``ItemListener'', and Tesseract also fails to correctly OCR ``ItemHandler''.
Google Vision API performs better, but it still has an OCR error (``s'' recognized as ``f'' due to the overlapping of the cursor over the ``s'' character).
For the bracket symbol, Tesseract recognizes it as a left parenthesis while Google Vision API misses this symbol.



OCR errors like missing brackets are relatively minor, but the erroneously OCRed identifiers like ``Icemflandlex'', ``ItemLiftener'' will affect the search and navigation of the source code in programming screencasts.
CodeTube~\cite{ponzanelli2016too} filters out these erroneously OCRed identifiers as noise from the OCRed text, but this post-processing may discard important code elements.
A better way is to correct as many OCR errors in the OCRed code as possible.
An intuition is that an erroneously OCRed identifier in one frame may be correctly recognized in another frame containing the same code.
For example, if the ``ItemListener'' in the next frame is not blocked by the cursor, it will be correctly recognized.
Therefore, the cross-frame information of the same code in the programming screenscat can help to correct OCR errors.
Another intuition is that we can learn a statistical language model from a corpus of source code and this language model can be used as a domain-specific spell checker to correct OCR errors in code.
In this work, we implement these two intuitions to fix errors in the OCRed code (see Section~\ref{sec:step4}), instead of simply ignoring them as noise.

\section{Approach}\label{sec:approach}
\begin{figure}[t]
    \centering
    \includegraphics[width=0.8\textwidth]{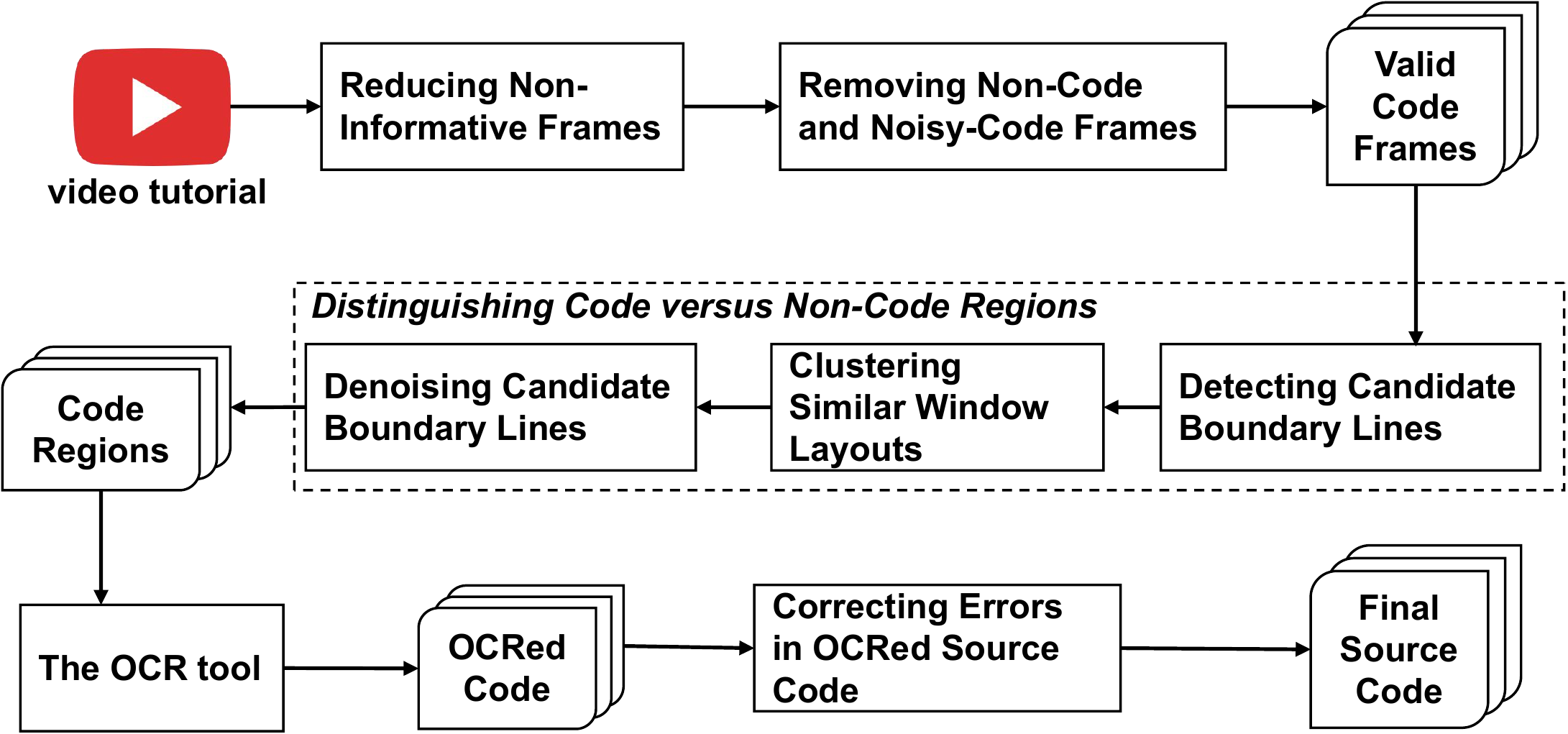}
    \caption{The work flow of \emph{psc2code}}\label{fig:process}
\end{figure}

Fig.~\ref{fig:process} describes the work flow of our \emph{psc2code} approach.
Given a programming screencast (e.g., YouTube programming video tutorial), \emph{psc2code} first computes the normalized root-mean-square error (NRMSE)
of consecutive frames and removes identical or almost-identical frames in the screencast.
Such identical or almost-identical frames are referred to as non-informative frames because analyzing them do not add new information to the extracted content.
Then, \emph{psc2code} leverages a CNN-based image classification technique to remove non-code frames (e.g., frames with Power Point slides) and noisy-code frames (e.g., frames with code completion suggestion popups).
Next, \emph{psc2code} detects boundaries of sub-windows in valid code frames and crops the frame regions that most likely contain code-editor windows.
In this step, \emph{psc2code} clusters close-by candidate boundary lines and frames with similar window layouts to reduce the noise for sub-window boundary detection.
Finally, \emph{psc2code} extracts the source code from the cropped code regions using the OCR technique and corrects the OCRed code based on cross-frame information in the screencast and a statistical language model of source code.

\subsection{Reducing Non-Informative Frames}\label{sec:step1}
A programming screencast recorded by the screencast tools usually contain a large portion of consecutive frames with no or minor differences, for example, when the developer does not interact with the computer or only moves the mouse or cursor.
There is no need to analyze each of such identical or almost-identical frames because they contain the same content.
Therefore, similar to existing programming video processing techniques~\cite{alahmadi2018accurately, ponzanelli2016too, ponzanelli2017automatic, moslehi2018feature}, the first step of \emph{psc2code} is to reduce such non-informative frames for further analysis.

Given a screencast, \emph{psc2code} first samples each second of the screencast.
It extracts the first frame of each second as an image using FFmpeg\footnote{http://www.ffmpeg.org/}.
We denote the sequence of the extracted frames as \{$f_i$\} ($1 \leq i \leq N$, $N$ being the last second of the screencast).
Then, it starts with the first extracted frame $f_1$ and uses an image differencing technique~\cite{wu2000spatial} to filter out subsequent frames with no or minor differences.
Given two frames $f_i$ and $f_j$ ($j \geq i+1$), \emph{psc2code} computes the normalized root-mean-square error (NRMSE) as the dissimilarity between the two frames, which is similar to the way used in the approach of Ponzanelli et al.~\cite{ponzanelli2016too}. NRMSE ranges from 0 (identical) to 1 (completely different).
If the dissimilarity between $f_i$ and $f_j$ is less than a user-specified threshold $T_{dissim}$ (0.05 in this work), \emph{psc2code} discards $f_j$ as a non-informative frame.
Otherwise, it keeps $f_i$ as an informative frame and uses $f_j$ as a new starting point to compare its subsequent frames.

\subsection{Removing Non-Code and Noisy-Code Frames}\label{sec:step2}
The goal of \emph{psc2code} is to extract code from frames in programming screencasts.
As discussed in Section~\ref{sec:motivation}, an informative frame may not contain code (see Fig.~\ref{fig:nocode} for a typical examples of non-code frames).
Furthermore, the code region of an IDE window in an informative frame may contain noise (e.g., code completion popups that block the real code).
Extracting content from such non-code and noisy-code frames not only wastes computing resources, but also introduces noise and hard-to-remove content irrelevant to the source code in the screencast.
Therefore, non-code and noisy-code frames have to be excluded from the subsequent code extraction steps.

The challenge in removing non-code and noisy-code frames lies in the fact that non-code and noisy-code frames vary greatly.
Non-code frames may involve many different software applications with diverse visual features (e.g., toolbar icons, sub-windows).
Furthermore, the window properties (e.g., size and position of sub-windows and popups) can be very different from one frame to another.
Such variations make it very complicated or even impossible to develop a set of comprehensive rules for removing non-code and noisy-code frames.

Inspired by the work of Ott et al.~\cite{OttAHBL08}, which trains a CNN-based classifier to classify programming languages of source code in programming videos, we design and train a CNN-based image classifier to identify non-code and noisy-code frames in programming screencasts.
Specifically, we formulate our task as a binary image classification problem, i.e., to predict whether a frame contains valid code or not:
\begin{itemize}
    \item Invalid frames: frames contain non-IDE windows or IDE windows with no or partially visible code.
    \item Valid code frames: frames contain IDE windows with at least an entire code editor window that contains completely-visible source code.
\end{itemize}
Instead of relying on human-engineered visual features to distinguish valid frames from invalid ones, the CNN model will automatically learn to extract important visual features from a set of training frames.

\subsubsection{Labeling Training Frames}\label{sec:training}
To build a reliable deep learning model, we need to label sufficient training data.
Although developers may use different system environments and tools in programming screencasts, the software applications and tools are used for the same purpose (e.g., code development, document editing, web browsing) and often share some common visual features.
Therefore, it is feasible to label a small amount of frames that contain typical software applications commonly used in programming screencasts for model training.

In this work, we randomly selected 50 videos from the dataset of programming screencasts we collected (see Table~\ref{tbl:playlists} for the summary of the dataset).
This dataset has 23 playlists of 1142 programming video tutorials from YouTube.
The 50 selected videos contain at least one video from each playlist.
Eight selected videos come from the playlists (P2, P12, P19 and P20) in which the developers do not use Eclipse as their IDE.
These 50 selected videos contain in total 5188 informative frames after removing non-informative ones following the steps in Section~\ref{sec:step1}.

To label these 5188 informative frames, we developed a web application that can show the informative frames of a programming screencast one by one.
Annotators can mark a frame as invalid frame or valid code frame by selecting a radio button.
Identifying whether a frame is a valid code frame or not is a straightforward task for human.
Ott et al. reported that a junior student usually spend 5 seconds to label an image~\cite{OttAHBL08}.
In this study, the first and second authors labeled the frames independently.
Both annotators are senior developers and have more than five years Java programming experience.
Each annotator spent approximately 10 hours to label all 5188 frames.
We use Fleiss Kappa\footnote{Fleiss Kappa of [0.01, 0.20], (0.20, 0.40], (0.40, 0.60], (0.60, 0.80] and (0.80, 1] is considered as slight, fair, moderate, substantial, and almost perfect agreement, respectively.}~\cite{fleiss1971measuring} to measure the agreement between the two annotators.
The Kappa value is 0.98, which indicates almost perfect agreement between the two annotators.
There are a small number of frames that the two annotators disagree with each other. For example, one annotator sometimes does not consider the frames with a tiny popup window such as the parameter hint when using functions (see Fig.~\ref{fig:annotatorerror}) as noisy-code frames.
For such frames, the two annotators discuss to determine the final label.
Table~\ref{tbl:label} presents the results of the labeled frames.
There are 1,864 invalid frames and 3,324 valid code frames, respectively.

\begin{figure}[t]
    \centering
    \includegraphics[width=0.48\textwidth]{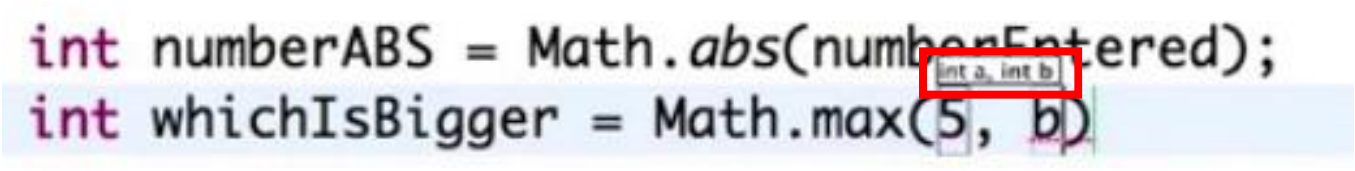}
    \caption{An example of annotator disagreement: a tiny popup (highlighted with the red rectangle)}\label{fig:annotatorerror}
\end{figure}

\begin{table}[]
    \centering
    \caption{Labeled frames used to train and test the CNN-based image classifier}\label{tbl:label}
    \begin{tabular}{@{}lll|l@{}}
    \toprule
             & Valid & Invalid & Total \\ \midrule
    Training & 2990     & 1679       & 4669  \\
    Testing  & 334      & 185        & 519   \\ \midrule
    Total    & 3324     & 1864       & 5188  \\ \bottomrule
    \end{tabular}
\end{table}

\subsubsection{Building the CNN-based Image Classifier}

We randomly divide the labeled frames into two parts: 90\% as the training data to train the CNN-based image classifier and 10\% as the testing data to evaluate the trained model.
We use 10-fold cross validation in model training and testing.
The CNN model requires the input images having a fixed size, but the video frames from different screencasts often have different resolutions.
Therefore, we rescale all frames to 300$\times$300 pixels.
We follow the approach used in the study of Ott et al.~\cite{OttAHBL08} that leverages a VGG network to predict whether a frame contains source code or not. A VGG network consists of multiple convolutional layers in succession followed by a max pooling layer for downsampling. It has been shown to perform well in identifying source code frames in programming screencast~\cite{OttAHBL08}.


We use {\tt Keras}\footnote{https://keras.io/} to implement our deep learning model. We set the maximum number of training iterations as 200.
We use accuracy as the metric to evaluate the trained model on the test data.
For the training of the CNN network we follow the appraoch of Ott et al.~\cite{OttAHBL08}, i.e., use the default trained VGG model in {\tt Keras} and only train the top layer of this model.
We run the model on a machine with Intel Core i7 CPU, 64GB memory, and one NVidia 1080Ti GPU with 16 GB of memory.
Finally, we obtain a CNN-based image classifier that achieves a score of 0.973 in accuracy on the testing data, which shows a very good performance on predicting whether a frame is a valid code frame or not.
We use this trained model to predict whether an unlabeled frame in our dataset of 1142 programming screencasts is a valid code frame or not.


\begin{figure*}[ht]
    \centering
    \includegraphics[width=0.95\textwidth]{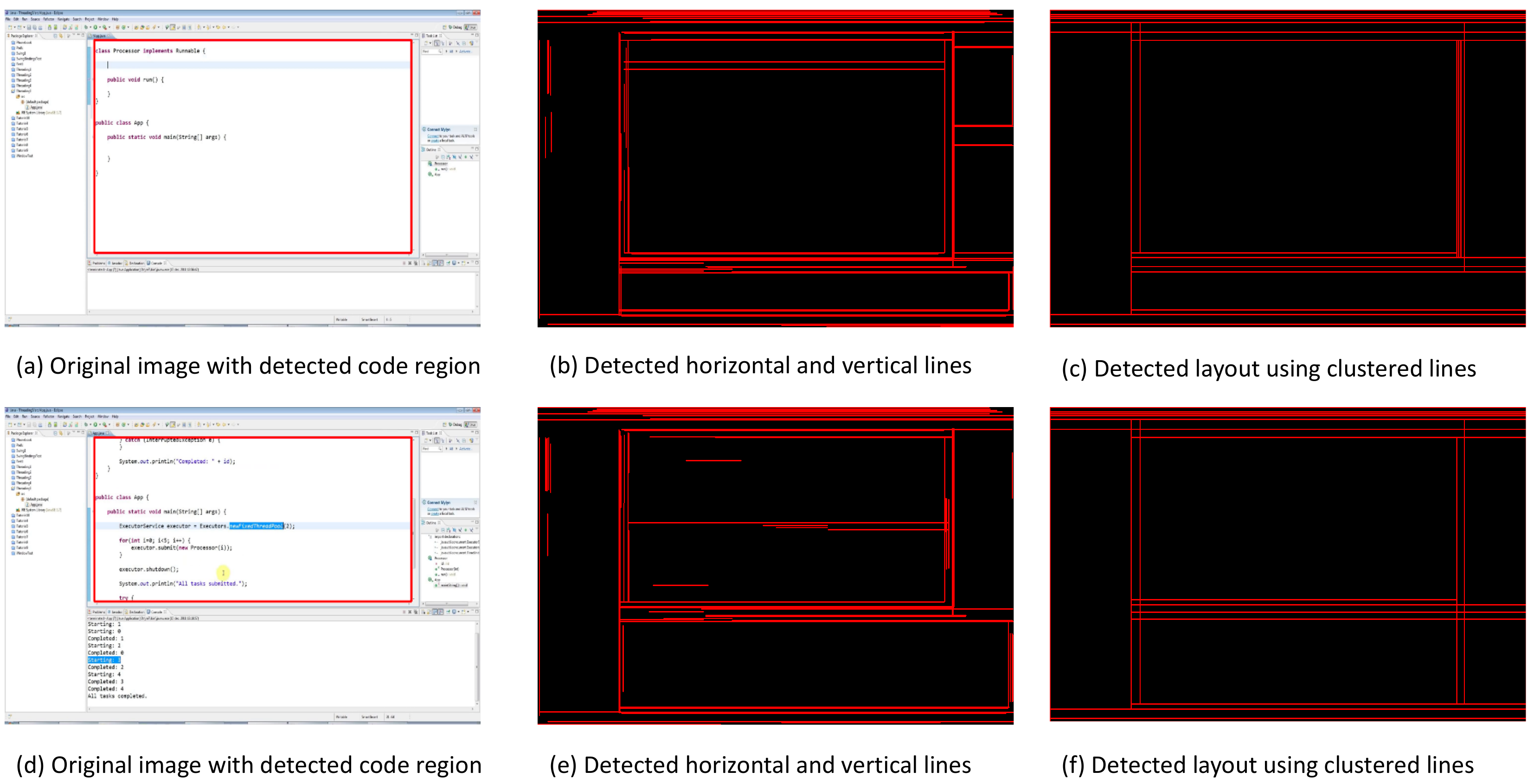}
    \caption{Illustration of sub-window boundary detection}\label{tbl:example}
\end{figure*}

\subsection{Distinguishing Code versus Non-Code Regions}\label{sec:step3}

A valid code frame predicted by the deep learning model should contain an entire non-blocked code editor sub-window in the IDE, but the frame usually contains many other parts of the IDE (e.g.,  navigation pane, outline view, console output, etc.) as well.
As discussed in Section~\ref{sec:motivation}, OCRing the entire frame will mix much noisy content from these non-code regions in the OCRed code from code regions.
A better solution is to crop the code region in a valid code frame and OCR only the code region.
As the sub-windows in the IDE have rectangle boundaries, an intuitive solution to crop the code region is to divide the frame into sub-windows by rectangle boundaries and then identify the sub-window that is most likely to be the code editor.
However, as shown in Fig.~\ref{tbl:example}, the unique characteristics of UI images often result in very noisy boundary detection results.
To crop the code region accurately, such noisy boundaries must be reduced.
Our \emph{psc2code} clusters close-by boundaries and similar window layout to achieve this goal.
It uses OpenCV APIs~\cite{opencv} for image processing.

\subsubsection{Detecting Candidate Boundary Lines}
We use the Canny edge detector~\cite{canny1986computational} to extract the edge map of a frame.
Probabilistic Hough transform~\cite{matas2000robust} is then used to get the horizontal and vertical lines, which are likely to be the boundaries of the sub-windows in the frame.
We filter the very short lines (less than 60 pixels in this work), which are unlikely to be sub-window boundaries.
Fig.~\ref{tbl:example} (b) and Fig.~\ref{tbl:example} (e) show the resulting horizontal and vertical lines for the frames in Fig.~\ref{tbl:example} (a) and Fig.~\ref{tbl:example} (d), respectively.
We can see that the detected horizontal and vertical lines are noisy.
To reduce noisy horizontal (or vertical) lines, we use the density-based clustering algorithm DBSCAN~\cite{ester1996density} to cluster the close-by horizontal (or vertical) lines based on their geometric distance and overlap.
Each line cluster is then represented by the longest line in the cluster.
By this step, we remove some close-by horizontal (or vertical) lines, which can reduce the complexity of sub-window boundary detection.


\subsubsection{Clustering Frames with Same Window Layouts}
Although clustering close-by lines reduce candidate boundary lines, there can still be many candidate boundary lines left which may complicate the detection of sub-window boundaries.
One observation we have for programming screencasts is that the developers do not frequently change the window layout during the recording of screencast.
For example, Fig.~\ref{tbl:example} (a) and Fig.~\ref{tbl:example} (b) show the two main window layouts in a programming video tutorial in our dataset of programming screencasts.
Fig.~\ref{tbl:example} (b) has a smaller code editor but a larger console output to inspect the execution results.
Note that the frames with the same window layout may have different horizontal and vertical lines, for example, due to presence/absence of code highlights, scrollbars, etc.
But the lines shared by the majority of the frames with the same layout are usually the boundaries of sub-windows.

To detect the frames with the same window layout, we cluster the frames based on detected horizontal and vertical lines in the frames.
Let $L=(h_1, h_2, ..., h_m, v_1, v_2, ..., v_n)$ be the set of the representative $m$ horizontal lines and $n$ vertical lines in a frame after clustering close-by lines.
Each line can then be assigned a unique index in $L$ and referred to as $L[i]$.
A frame $f$ can be denoted as a line vector $V(f)$, which is defined as follows: $$V(f)=(ind(L[0], f), ..., ind(L[m+n], f))$$ where $ind(l, f)=1$ if the frame $f$ contains the line $l$, and $ind(l, f)=0$ otherwise.
We then use the DBSCAN clustering algorithm to cluster the frames in a programming screencast based on the distance between their line vectors.
This step results in some clusters of frames, each of which represents a distinct window layout.
For each cluster of frames, we keep only the lines shared by the majority frames in the cluster.
In this way, we can remove noisy candidate boundary lines that appears in only some frames but not others.
Fig.~\ref{tbl:example} (c) and Fig.~\ref{tbl:example} (f) show the resulting boundary lines based on the analysis of the common lines in the frames with the same window layout.
We can see that many noisy lines such as those from different code highlights are successfully removed.


\subsubsection{Detecting Sub-Windows and Code Regions}
Based on the clear boundary lines after the above two steps of denoising candidate boundary lines,
we detect sub-windows by forming rectangles with the boundary lines.
When several candidate rectangles overlap, we keep the smallest rectangle as the sub-window boundary.
This allows us to crop the main content region of the sub-windows but ignore window decorators such as scrollbars, headers and/or rulers.
We observe that code editor window in the IDE usually occupies the largest area in the programming screencasts in our dataset.
Therefore, we select the detected sub-window with the largest rectangle area as the code region.
The detected code regions for the two frames in Fig.~\ref{tbl:example} (a) and Fig.~\ref{tbl:example} (d) are highlighted in red box in the figures.
Note that one can also develop image classification method such as the method proposed in Section~\ref{sec:step2} to distinguish code-editor sub-windows from non-code-editor sub-windows. However, we take a simpler heuristic-based method in this work, because it is more efficient than deep learning model and is sufficient for our experimental screencasts.

\begin{figure}
    \includegraphics[width=0.7\textwidth]{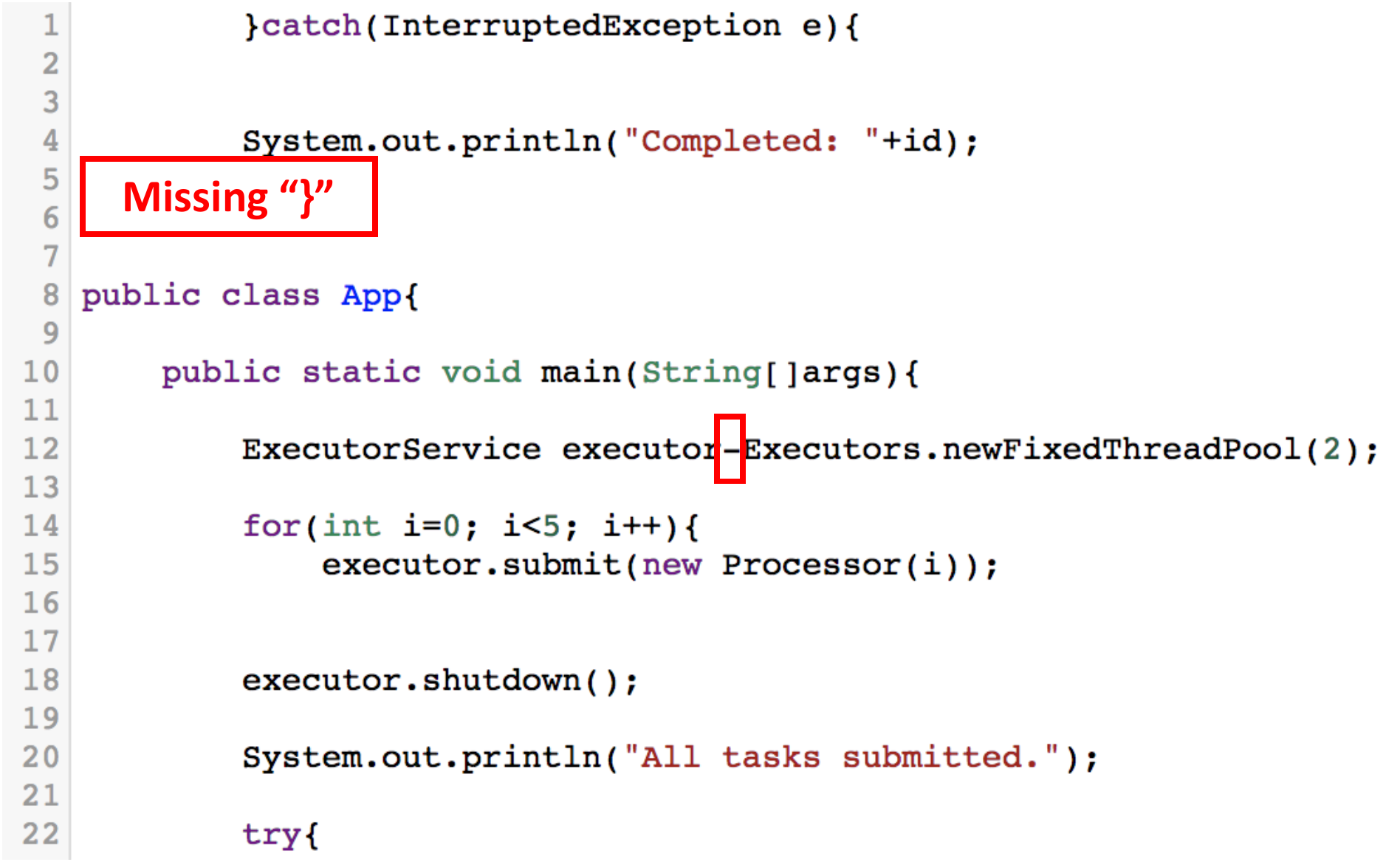}
    \caption{The OCRed source code for the code region in Fig.~\ref{tbl:example} (d); the OCR errors are highlighted by the red rectangles.}\label{tbl:ocr}
\end{figure}

\subsection{Correcting Errors in OCRed Source Code}\label{sec:step4}
Given an image of the cropped code region, we use the Google Vision API~\cite{googlevision} to extract source code for the image.
The Google Vision API for text detection returns the OCR result in the format of JSON, which includes the entire extracted string, as well as individual words and their bounding boxes.
We can reconstruct the extracted string into the formatted source code based on the position of words.
Fig.~\ref{tbl:ocr} shows the OCRed source code for the code region in the frame in Fig~\ref{tbl:example} (d).
We can see that there are some OCR errors.
For example, some brackets are missing (Line 5 and 6); in Line 12, the symbol `=' is recognized as `-'.
Furthermore, we often observe that the cursor is recognized as `i' or 'I', which results in incorrect words.

Many existing techniques~\cite{ponzanelli2016too, khandwala2018codemotion}, except Yadid and Yahav~\cite{yadid2016extracting}, simply discard the words with OCR errors. In contrast, we survey the literature and choose to integrate the heuristics in the two previous works~\cite{yadid2016extracting, khandwala2018codemotion} to fix as many OCR errors as possible.
First, we use the effective heuristics of Kandarp and Guo~\cite{khandwala2018codemotion} to remove line numbers and fix Unicode errors:

\begin{itemize}[leftmargin=*]
    \item Sometimes there are line numbers displayed on the left edge of the cropped image. To deal with them, we first identify whether there are numbers at the begin of the lines. If yes, we remove these numbers in the OCRed lines.

    \item Due to image noise, the OCR technique sometimes erroneously recognizes text within images as accented characters (e.g., \`{o} or \'{o} ) or Unicode variants. We convert these characters into their closest unaccented ASCII versions.
\end{itemize}

Then, we integrate the approach of Yadid and Yahav~\cite{yadid2016extracting} to further correct the OCR errors.
This approach assumes that an erroneously OCRed word in one frame may be correctly recognized in another frame containing the same code.
Take the OCR errors in Fig.~\ref{fig:lineocr} as an example.
If the ``ItemListener'' in the next frame is not blocked by the cursor, it will be correctly recognized.
Therefore, the cross-frame information of the same code can help to correct OCR errors.
Furthermore, this approach learns a statistical language model from a corpus of source code as a domain-specific spell checker to correct OCR errors.
\begin{itemize}[leftmargin=*]
\item For some incorrect words or lines of code, we can find the correct one in other related frames. Thus, we follow the approach of Yadid and Yahav~\cite{yadid2016extracting}, which uses cross-frame information and statistical language models to correct these minor errors in the OCRed text.
    First, we use the top 300 starred Github Java projects to build a unigram language model and a line structure model, which captures frequent \textit{line structures} permitted by the grammar of the language. We build the unigram language model based on the extracted tokens from source code and build the line structure model based on the common line structures using token types.
    Second, given the OCRed source code extracted from a video, we detect the incorrect words and lines of code based on the statistical language models.
    Finally, given an incorrect line of code (or word), we find the most similar correct line (or word) with sufficient confidence based on edit distance to correct it.

    To illustrate this process, we use an example line of code from our dataset: {\tt Properties propIn = new Properties();}, whose line structure is denoted as {\tt IDU IDL = new IDU ( ) ;}, where {\tt IDU} is the identifier starting with upper character and {\tt IDL} is the identifier starting with lower character. An incorrect OCR text of this line from a frame is {\tt Properties prop In - new Properties();}. There is an extraneous space between {\tt prop} and {\tt in} and the symbol `=' is recognized as `-'. Thus, its line structure become {\tt IDU IDL IDU - new IDU ( ) ;}, which is likely to be incorrect based on our constructed statistical line structure model.
    To make correction for this line, we find the correct line detected based on edit distance in another frame.

\end{itemize}

\section{Experiments}\label{sec:experiment}
\subsection{Experiment Setup}
\subsubsection{Programming Video Tutorial Dataset}\label{sec:dataset}
\begin{table*}[]
    \centering
    \caption{The YouTube playlists used in the study. ID=the index of a playlist, Playlist Name=the title of the playlist in YouTube, \#Videos=number of videos used in the study, Average Dur. (Sec)=average video duration in seconds, Res.=resolution of the videos in a playlist, Average \#Informative=average number of informative frames, Average \#Valid=average number of valid code frames, Average \#Valid/\#Informative=ratio of valid code frames out of informative frames.}
    \label{tbl:playlists}
    \resizebox{\textwidth}{!}{%
    \begin{tabular}{@{}llrrrrrr@{}}
    \toprule
    ID  & Playlist Name                                       & \#Videos & \begin{tabular}[c]{@{}c@{}}Average \\Dur. (Sec)\\\end{tabular} & Res. & \begin{tabular}[c]{@{}c@{}} Average\\ \#Informative \\\end{tabular} & \begin{tabular}[c]{@{}c@{}}Average \\ \#Valid \\ \end{tabular} & \begin{tabular}[c]{@{}c@{}}Average \\ \#Valid/\#Informative\\ \end{tabular}\\  \midrule
    P1  & \href{https://www.youtube.com/playlist?list=PL27BCE863B6A864E3}{Java (Intermediate) Tutorials}                       & 59   & 362 & 360p &  63 & 32 & 0.51  \\
    P2  & \href{https://www.youtube.com/playlist?list=PL4unWLKFsZfeMAFF5NEfrosw7Z-UQVN90}{Java Tutorial in Tamil}                              & 56  & 369 & 720p & 83 & 45 & 0.54   \\
    P3  & \href{https://www.youtube.com/playlist?list=PL6n9fhu94yhWizLudXueXf16yJzlwNrSc}{Java tutorial for beginners}                         & 20  & 819 & 720p & 89 & 50 & 0.56   \\
    P4  & \href{https://www.youtube.com/playlist?list=PL71C6DFDDF73835C2}{Java Tutorials}                                      & 98  &  421 & 720p &  65 & 43 & 0.66   \\
    P5  & \href{https://www.youtube.com/playlist?list=PL9ooVrP1hQOFR25JoQW3h3n5pBs77e6KU}{Java Online Training Videos}                         & 25   & 4,067 & 720p & 335 & 216 & 0.64    \\
    P6  & \href{https://www.youtube.com/playlist?list=PLBB24CFB073F1048E}{Java Multithreading}                                 & 14  & 686 & 720p  & 100 & 80 & 0.80  \\
    P7  & \href{https://www.youtube.com/playlist?list=PLCZlgfAG0GXDUvrO3Bc_VUvIjWKnYIRJ1}{Belajar Java Untuk Pemula}                           & 44  & 243  & 720p & 28 & 19 & 0.68   \\
    P8  & \href{https://www.youtube.com/playlist?list=PLE7E8B7F4856C9B19}{Java Video Tutorial}                                 & 90   & 852 & 720p &  161 & 75 & 0.47  \\
    P9  & \href{https://www.youtube.com/playlist?list=PLF03C6B2C0B292A1E}{Java Programming with Eclipse Tutorials}             & 60 &  576 & 720p & 78 & 53 & 0.68   \\
    P10 & \href{https://www.youtube.com/playlist?list=PLFE2CE09D83EE3E28}{Java (Beginner) Programming Tutorials}               & 84 & 420 & 720p & 85 & 54 & 0.64    \\
    P11 & \href{https://www.youtube.com/playlist?list=PLGeQyNDhU6x3GC-WGSYR6bfdX7O76wR_F}{Tutorial Java}                                       & 52 & 444 & 360p & 50 & 35 & 0.70    \\
    P12 & \href{https://www.youtube.com/playlist?list=PLOJzCFLZdG4zk5d-1_ah2B4kqZSeIlWtt}{Java Chess Engine Tutorial}                          & 52  & 928 & 720p & 138 & 101 & 0.73  \\
    P13 & \href{https://www.youtube.com/playlist?list=PLOUYE-KsFYc_0ekC_3EEAIkiRuEIaymmJ}{Advanced Java tutorial}                              & 58  & 297 & 720p & 40 & 25 & 0.63   \\
    P14 & \href{https://www.youtube.com/playlist?list=PLS1QulWo1RIbfTjQvTdj8Y6yyq4R7g-Al}{Java Tutorial For Beginners (Step by Step tutorial)} & 72  & 597 & 720p & 89 & 54 & 0.61   \\
    P15 & \href{https://www.youtube.com/playlist?list=PLah6faXAgguMnTBs3JnEJY0shAc18XYQZ}{NEW Beginner 2D Game Programming}                    & 33 &  716 & 720p & 183 & 123 & 0.67   \\
    P16 & \href{https://www.youtube.com/playlist?list=PLbu9W4c-C0iBjwSxY_zFrNQG1tarm4K7T}{Socket Programming in Java}                          & 4  & 472 & 720p & 42 & 18 & 0.43    \\
    P17 & \href{https://www.youtube.com/playlist?list=PLqq-6Pq4lTTZh5U8RbdXq0WaYvZBz2rbn}{Developing RESTful APIs with JAX-RS}                 & 18 & 716 & 720p & 153 & 66 & 0.43    \\
    P18 & \href{https://www.youtube.com/playlist?list=PLqq-6Pq4lTTa9YGfyhyW2CqdtW9RtY-I3}{Java 8 Lambda Basics}                                & 18 & 488 & 720p & 62 & 35 & 0.56    \\
    P19 & \href{https://www.youtube.com/playlist?list=PLr6-GrHUlVf9SIx5cDhoEMknias5Xyv67}{Java Tutorial for Beginners}                         & 55 & 348 & 720p & 40 & 32 & 0.80    \\
    P20 & \href{https://www.youtube.com/playlist?list=PLsyeobzWxl7oZ-fxDYkOToURHhMuWD1BK}{Java Tutorial For Beginners}                         & 60 & 312 & 720p & 38 & 31 & 0.82    \\
    P21 & \href{https://www.youtube.com/playlist?list=PLsyeobzWxl7pFZoGT1NbZJpywedeyzyaf}{Java Tutorial for Beginners 2018}                    & 56 & 363 & 720p & 63 & 46 & 0.73    \\
    P22 & \href{https://www.youtube.com/playlist?list=PLv6UtFrA7VEu4PtzJaGHHSeZBi6mdJtwv}{Eclipse and Java for Total Beginners}                & 16   & 723 & 360p & 159 & 62 & 0.39    \\
    P23 & \href{https://www.youtube.com/playlist?list=PLzS3AYzXBoj8Fm19MdV3jqZOCLOh984xK}{Java 8 Features Tutorial(All In One)}                & 98 & 622 & 720p & 113 & 75 & 0.66   \\ \midrule
    Average & & 50 & 593 & & 91 & 56 & 0.62 \\ \bottomrule
    \end{tabular}
    }
\end{table*}

In this study, our targeted programming screencasts are live coding video tutorials where tutorial authors demonstrate how to write a program in IDEs (e.g., Eclipse, Intellij).
We focus on Java programming in this study but it is easy to extend our approach to other programming languages.
Since YouTube has a large number of programming video tutorials and also provide YouTube Data APIs\footnote{https://developers.google.com/youtube/v3/} to access and search videos easily, we build our dataset of programming video tutorials based on YouTube videos.

We used the query ``Java tutorial'' to search video playlists using YouTube Data API. From the search results, we considered the top-50 YouTube playlists ranked by the playlists' popularity.
However, we did not use all these 50 playlists because some tutorial authors did not live coding in IDEs but use other tools (e.g., Power Point slides) to explain programming concepts and code, or the videos in the playlist are not screencast videos.
Finally, we used 23 playlists in this study, which are listed in Table~\ref{tbl:playlists}.
For each playlist, we downloaded its videos at the maximum available resolution and the corresponding audio transcripts using {\tt pytube}\footnote{https://github.com/nficano/pytube}. 
We further found that not all downloaded videos include writing and editing source code.
For example, some video tutorials just introduce how to install JDK and IDEs. 
We removed such videos and got in total 1,142 videos as our dataset for the experiments\footnote{All videos can be found: \url{https://github.com/baolingfeng/psc2code}}.

As shown in Table~\ref{tbl:playlists}, our dataset is diverse in terms of programming knowledge covered, development tools used and video statistics.
Many playlists are for Java beginners. But there are several playlists that provide advanced Java programming knowledge, such as multithreading (P6), chess game (P12), and 2D game programming (P15).
Among the 23 playlists, the majority of tutorial authors use Eclipse as their IDE, while some tutorial authors use other IDEs including NetBeans (P19, P20), Intellij IDEA (P12), and Notepad++ (P2). 
The duration of most videos is 5 to 10 minutes except those in the playlist P5 that have the duration of more than one hour.
This is because each video in P5 covers many concepts while other playlists usually cover only one main concept in one video.
Most of videos have the resolution of 720p (1280$\times$720) except for the videos in the playlist P1, P11 and P22 that have the resolution 360p (480$\times$360).

We applied \emph{psc2code} to extract source code from these 1,142 programming videos. 
After performing the step of reducing redundant frames (Section~\ref{sec:step1}), the number of informative frames left is not very large. 
For example, the videos in the playlist P5 are of long duration but there are only 335 informative frames left per video on average. 
After applying the CNN-based image classifier to remove non-code and noisy-code frames (Section~\ref{sec:step2}), more frames are removed.
As shown in Table~\ref{tbl:playlists}, about 62\% of informative frames are identified as valid code frames on average across the 23 playlists.
\emph{psc2code} identifies code regions in these valid code frames, and finally OCRs the source code from the code regions and corrects the OCRed source code.

\subsubsection{Research Questions}

We conduct a set of experiments to evaluate the performance of \emph{psc2code}, aiming to answer the following research questions:




\vspace{0.1cm} \noindent  \textbf{RQ1: Can our approach effectively remove non-informative frames?}

\vspace{0.1cm} \noindent \textbf{Motivation.}
The first step of our approach removes a large a large portion of consecutive frames with no or minor differences, which are considered as non-informative frames. In this research question, we want to investigate whether these removed frames are truly non-informative.

\vspace{0.1cm} \noindent \textbf{RQ2: Can the trained CNN-based image classifier effectively identify the non-code and noisy-code frames in the unseen programming screencasts?}

\vspace{0.1cm} \noindent \textbf{Motivation.}
To train the CNN-based image classifier for identifying non-code and noisy-code frames, we select and label a small subset of programming videos from our dataset.
Although the trained model achieves 97\% accuracy on the testing data (See Section~\ref{sec:training}), the performance of the trained model on the remaining programming videos that are completely unseen during model training needs to be evaluated.
If the trained model cannot effectively identify the non-code and noisy-code frames in these unseen videos, it will affect the subsequent processing steps.



\vspace{0.1cm} \noindent \textbf{RQ3: Can our approach accurately locate code regions in code frames?}

\vspace{0.1cm} \noindent \textbf{Motivation.}
To detect code regions in code frames, our approach leverages computer vision techniques to identify candidate boundary lines and cluster frames with same window layouts. However, due to the noise (e.g., highlighted lines in IDEs) in these frames, our approach might detect incorrect code regions. So, we want to investigate whether our approach can accurately locate code regions in code frames.

\vspace{0.1cm} \noindent \textbf{RQ4: Can our approach effectively fix the OCR errors in the source code OCRed from programming videos?}

\vspace{0.1cm} \noindent \textbf{Motivation.}
There are still many OCR errors in the OCRed source code even when we use a professional OCR tool. Our \emph{psc2code} uses the cross-frame information in the screencast and a statistical language source code model to correct these errors.
We want to investigate whether the corrections made by our approach are truly correct and improve the quality of the OCRed source code.

\vspace{0.1cm} \noindent \textbf{RQ5: Can our approach efficiently extract and denoise code from programming screencast?}

\vspace{0.1cm} \noindent \textbf{Motivation.}
In this research question, we want to investigate the scalability of our approach. We want to know how long each step in our approach take, such as extracting the informative frames, locating code regions, and correcting the extracted code fragments. 

\subsection{Experiment Results}

\subsubsection{Effectiveness of Removing Non-Informative Frames (RQ1)}\label{sec:rq1}

\vspace{0.1cm} \noindent \textbf{Approach.}
Since the number of the removed non-informative frames is too large, it is impossible to verify all the non-informative frames. Therefore, given a video, we randomly sample one frame between the two adjacent frames keeped in the first step if the interval of these two frames is larger than 1 second. 
We randomly sample one video from each playlist, and we obtain 1189 non-informative frames in total. 
Because we are interested in source code in this study, the annotators label a discarded frame as truly non-informative if a completed statement in its source code can be found in the corresponding position of the kept frames (i.e., informative frames), otherwise we regard the frame as an informative frame, which is incorrectly discarded.
The two annotators who label the frames for the training data verify the sampled non-informative frames. 

\medskip \noindent \textbf{Results.}
We find that there are 11 frames out of the 1189 non-informative frames (less than 1\%), which contains at least one completed statement that cannot be found in the informative frames. These 11 frames are from 6 videos, and the largest number of incorrect non-informative frames for a video is 3. 
However, we find that the source code in these discarded informative frames are usually intermediate and will be changed by the developers in a short time. For example, in the sampled video of the playlist P1, there are three informative frames found in the non-informative frames. One of the three informative frame has three same statements {\tt System.out.println("Sophomore")} with different {\tt if} condition, which is generated by copy and paste. While among the informative frames we only find that there are three {\tt System.out.println} statement with {\tt "Sophomore", "Junior", "Senior"} in some frames, respectively. 
Overall, we think the information loss is small and would not affect our final analysis results, because only a small proportion of unimportant information is dismissed. 


\subsubsection{Effectiveness of Identifying Non-Code and Noisy-Code Frames (RQ2)}\label{sec:rq2}

\vspace{0.1cm} \noindent \textbf{Approach.}
Except the 50 programming videos labeled for model training, we still have more than 1,000 unlabeled videos with a large number of informative frames.
It is difficult to manually verify the classification results of all these informative frames.
Therefore, to verify the performance of the trained CNN-based image classifier,
we randomly sample two videos from each playlist, which provided us with 46 videos with 4828 informative frames in total.
Examining the classification results of these sampled frames can give us the accuracy metrics at the 95\% confidence level with an error margin of 1.38\%.


There are 1924 invalid frames and 2904 valid code frames as predicted by the trained image classifier, respectively (see Table~\ref{tbl:rq1}).
The ratio of invalid frames versus valid code frames in the examined 46 videos is similar to that of the 50 programming videos for model training (see Table~\ref{tbl:label}).
The two annotators who label the frames for the training data verify the classification results of the 4828 frames in the 46 videos.
We also use Fleiss Kappa to measure the agreement between the two annotators.
The Kappa value is 0.97, which indicates almost perfect agreement between the two annotators.
For the classification results that the two annotators disagree, they discuss to reach an agreement.

For each frame, there can be 4 possible outcomes predicted by the image classifier: a valid code frame is classified as valid (true positive TP); an invalid frame is classified as valid (false positive FP); a valid code frame is classified as invalid (false negative FN); an invalid frame is classified as invalid (true negative TN).
For each playlist, we construct a confusion matrix based on these four possible outcomes, then calculate the accuracy, precision, recall and F1-score.
We also merge all the confusion matrixes of all playlists and compute the overall metrics for the performance of the image classifier.

\begin{itemize}[leftmargin=*]
    \item \textbf{Accuracy}: the number of correctly classified frames (both valid and invalid) over the total number of frames, i.e, $Acc=\frac{TP+TN}{TP+FP+FN+TN}$.
    \item \textbf{Precision on valid frames}: the proportion of frames that are    correctly predicted as valid among all frames predicted as valid, i.e., $P(V)=\frac{TP}{TP+FP}$.
    \item \textbf{Recall on valid frames}: the proportion of valid frames that are correctly predicted, i.e., $R(V)=\frac{TP}{TP+FN}$.
    \item \textbf{Precision on invalid frames}: the proportion of frames that are correctly predicted as invalid among all frames predicted as invalid, i.e., $P(IV)=\frac{TN}{TN+FN}$.
    \item \textbf{Recall on invalid frames}: the proportion of invalid frames that are correctly predicted, i.e., $R(IV)=\frac{TN}{TN+FP}$.
    \item \textbf{F1-score}: a summary measure that combines both precision and recall - it evaluates if an increase in precision (recall) outweighs a reduction in recall (precision). For F1-score of valid frames, it is $F(V)=\frac{2\times P(V)\times R(V)}{P(V)+R(V)}$. For F1-score of invalid frames, it is $F(IV)=\frac{2\times P(IV)\times R(IV)}{P(IV)+R(IV)}$.
\end{itemize}

\begin{table*}[]
    \centering
    \caption{The results of distinguishing invalid frames from valid code frames by the CNN-based image classifier of \emph{psc2code} }\label{tbl:rq1}
    \resizebox{\textwidth}{!}{%
    \begin{tabular}{@{}l|rrrrrr|rrrrrrr@{}}
        \toprule
        \multirow{2}{*}{Playlist} & \multirow{2}{*}{\#Valid} & \multirow{2}{*}{\#Invalid} & \multirow{2}{*}{TP} & \multirow{2}{*}{FP} & \multirow{2}{*}{TN} & \multirow{2}{*}{FN} & \multirow{2}{*}{Accuracy} & \multicolumn{3}{c}{Valid frames} & \multicolumn{3}{c}{Invalid frames} \\
                                  &                              &                                &                     &                     &                     &                     &                           & Precision    & Recall   & F1-score   & Precision    & Recall    & F1-score    \\ \midrule
                                  P1  & 122     & 40        & 121  & 7   & 33   & 1   & 0.95     & 0.95       & 0.99    & 0.97       & 0.97      & 0.83    & 0.89       \\
                                  P2  & 108     & 102       & 100  & 11  & 91   & 8   & 0.91     & 0.90       & 0.93    & 0.91       & 0.92      & 0.89    & 0.91       \\
                                  P3  & 91      & 73        & 88   & 18  & 55   & 3   & 0.87     & 0.83       & 0.97    & 0.89       & 0.95      & 0.75    & 0.84       \\
                                  P4  & 104     & 26        & 64   & 1   & 25   & 40  & 0.68     & 0.98       & 0.62    & 0.76       & 0.38      & 0.96    & 0.55       \\
                                  P5  & 448     & 358       & 362  & 28  & 330  & 86  & 0.86     & 0.93       & 0.81    & 0.86       & 0.79      & 0.92    & 0.85       \\
                                  P6  & 117     & 37        & 115  & 8   & 29   & 2   & 0.94     & 0.93       & 0.98    & 0.96       & 0.94      & 0.78    & 0.85       \\
                                  P7  & 93      & 49        & 93   & 12  & 37   & 0   & 0.92     & 0.89       & 1.00    & 0.94       & 1.00      & 0.76    & 0.86       \\
                                  P8  & 154     & 91        & 149  & 7   & 84   & 5   & 0.95     & 0.96       & 0.97    & 0.96       & 0.94      & 0.92    & 0.93       \\
                                  P9  & 103     & 18        & 75   & 1   & 17   & 28  & 0.76     & 0.99       & 0.73    & 0.84       & 0.38      & 0.94    & 0.54       \\
                                  P10 & 117     & 29        & 114  & 2   & 27   & 3   & 0.97     & 0.98       & 0.97    & 0.98       & 0.90      & 0.93    & 0.92       \\
                                  P11 & 99      & 39        & 13   & 0   & 39   & 86  & 0.38     & 1.00       & 0.13    & 0.23       & 0.31      & 1.00    & 0.48       \\
                                  P12 & 242     & 111       & 201  & 15  & 96   & 41  & 0.84     & 0.93       & 0.83    & 0.88       & 0.70      & 0.86    & 0.77       \\
                                  P13 & 35      & 24        & 32   & 6   & 18   & 3   & 0.85     & 0.84       & 0.91    & 0.88       & 0.86      & 0.75    & 0.80       \\
                                  P14 & 87      & 53        & 76   & 7   & 46   & 11  & 0.87     & 0.92       & 0.87    & 0.89       & 0.81      & 0.87    & 0.84       \\
                                  P15 & 324     & 109       & 295  & 15  & 94   & 29  & 0.90     & 0.95       & 0.91    & 0.93       & 0.76      & 0.86    & 0.81       \\
                                  P16 & 21      & 64        & 19   & 1   & 63   & 2   & 0.96     & 0.95       & 0.90    & 0.93       & 0.97      & 0.98    & 0.98       \\
                                  P17 & 98      & 197       & 98   & 51  & 146  & 0   & 0.83     & 0.66       & 1.00    & 0.79       & 1.00      & 0.74    & 0.85       \\
                                  P18 & 69      & 92        & 54   & 7   & 85   & 15  & 0.86     & 0.89       & 0.78    & 0.83       & 0.85      & 0.92    & 0.89       \\
                                  P19 & 70      & 81        & 69   & 5   & 76   & 1   & 0.96     & 0.93       & 0.99    & 0.96       & 0.99      & 0.94    & 0.96       \\
                                  P20 & 69      & 52        & 62   & 1   & 51   & 7   & 0.93     & 0.98       & 0.90    & 0.94       & 0.88      & 0.98    & 0.93       \\
                                  P21 & 77      & 22        & 74   & 3   & 19   & 3   & 0.94     & 0.96       & 0.96    & 0.96       & 0.86      & 0.86    & 0.86       \\
                                  P22 & 135     & 156       & 102  & 39  & 117  & 33  & 0.75     & 0.72       & 0.76    & 0.74       & 0.78      & 0.75    & 0.76       \\
                                  P23 & 121     & 101       & 83   & 11  & 90   & 38  & 0.78     & 0.88       & 0.69    & 0.77       & 0.70      & 0.89    & 0.79       \\ \midrule
                                  All & 2904    & 1924      & 2459 & 256 & 1668 & 445 & 0.85     & 0.91       & 0.85    & 0.88       & 0.79      & 0.87    & 0.83       \\ \bottomrule
                                       \end{tabular}
    }
\end{table*}

\begin{table}[]
    \centering
    \caption{The comparison results between our approach and the baseline.}\label{tbl:comparison}
    \resizebox{!}{.3\paperheight}{%
    \begin{tabular}{@{}llrrrr@{}}
    \toprule
    Playlist & Approach & Accuracy & F1-score@valid & F1-score@invalid & IOU  \\ \midrule
    P1       & Ours     & 0.95     & 0.97           & 0.89             & 0.78 \\
             & Baseline & 0.62     & 0.66           & 0.56             & 0.49 \\ \midrule
    P2       & Ours     & 0.91     & 0.91           & 0.91             & 0.97 \\
             & Baseline & 0.73     & 0.71           & 0.75             & 0.67 \\ \midrule
    P3       & Ours     & 0.87     & 0.89           & 0.84             & 0.98 \\
             & Baseline & 0.91     & 0.92           & 0.90             & 0.77 \\ \midrule
    P4       & Ours     & 0.68     & 0.76           & 0.55             & 1.00 \\
             & Baseline & 0.73     & 0.80           & 0.59             & 0.66 \\ \midrule
    P5       & Ours     & 0.86     & 0.86           & 0.85             & 0.86 \\
             & Baseline & 0.66     & 0.60           & 0.70             & 0.59 \\ \midrule
    P6       & Ours     & 0.94     & 0.96           & 0.85             & 0.89 \\
             & Baseline & 0.76     & 0.82           & 0.62             & 0.62 \\ \midrule
    P7       & Ours     & 0.92     & 0.94           & 0.86             & 0.92 \\
             & Baseline & 0.96     & 0.97           & 0.94             & 0.84 \\ \midrule
    P8       & Ours     & 0.95     & 0.96           & 0.93             & 0.97 \\
             & Baseline & 0.94     & 0.95           & 0.92             & 0.79 \\ \midrule
    P9       & Ours     & 0.76     & 0.84           & 0.54             & 0.95 \\
             & Baseline & 0.76     & 0.84           & 0.48             & 0.59 \\ \midrule
    P10      & Ours     & 0.97     & 0.98           & 0.92             & 0.94 \\
             & Baseline & 0.62     & 0.70           & 0.48             & 0.50 \\ \midrule
    P11      & Ours     & 0.38     & 0.23           & 0.48             & 0.90 \\
             & Baseline & 0.38     & 0.28           & 0.45             & 0.35 \\ \midrule
    P12      & Ours     & 0.84     & 0.88           & 0.77             & 0.94 \\
             & Baseline & 0.75     & 0.80           & 0.68             & 0.61 \\ \midrule
    P13      & Ours     & 0.85     & 0.88           & 0.80             & 0.89 \\
             & Baseline & 0.87     & 0.89           & 0.84             & 0.74 \\ \midrule
    P14      & Ours     & 0.87     & 0.89           & 0.84             & 0.94 \\
             & Baseline & 0.49     & 0.29           & 0.60             & 0.47 \\ \midrule
    P15      & Ours     & 0.90     & 0.93           & 0.81             & 0.98 \\
             & Baseline & 0.72     & 0.78           & 0.60             & 0.60 \\ \midrule
    P16      & Ours     & 0.96     & 0.93           & 0.98             & 0.93 \\
             & Baseline & 0.90     & 0.82           & 0.93             & 0.86 \\ \midrule
    P17      & Ours     & 0.83     & 0.79           & 0.85             & 0.90 \\
             & Baseline & 0.72     & 0.61           & 0.78             & 0.67 \\ \midrule
    P18      & Ours     & 0.86     & 0.83           & 0.89             & 0.95 \\
             & Baseline & 0.90     & 0.88           & 0.91             & 0.79 \\ \midrule
    P19      & Ours     & 0.96     & 0.96           & 0.96             & 0.85 \\
             & Baseline & 0.85     & 0.84           & 0.86             & 0.73 \\ \midrule
    P20      & Ours     & 0.93     & 0.94           & 0.93             & 0.85 \\
             & Baseline & 0.81     & 0.85           & 0.75             & 0.70 \\ \midrule
    P21      & Ours     & 0.94     & 0.96           & 0.86             & 0.92 \\
             & Baseline & 0.63     & 0.69           & 0.53             & 0.47 \\ \midrule
    P22      & Ours     & 0.75     & 0.74           & 0.76             & 0.91 \\
             & Baseline & 0.72     & 0.61           & 0.78             & 0.67 \\ \midrule
    P23      & Ours     & 0.78     & 0.77           & 0.79             & 0.87 \\
             & Baseline & 0.73     & 0.72           & 0.74             & 0.62 \\ \midrule
    All      & Ours     & 0.85     & 0.88           & 0.83             & 0.92 \\
             & Baseline & 0.73     & 0.74           & 0.72             & 0.64 \\ \bottomrule
    \end{tabular}
    }
\end{table}

\medskip \noindent \textbf{Results.}
Table~\ref{tbl:rq1} presents the results of the prediction performance for the 4828 frames in the 46 examined programming videos (in the last row, the first 6th columns are the sum of all playlists, and the last seven columns are the overall metrics).
The overall accuracy is 0.85, and the overall F1-score on invalid frames and valid code frames are 0.88 and 0.83, respectively.
These results indicate that the image classifier of \emph{psc2code} can idenify valid code frames from invalid ones effectively.

In terms of accuracy, is often larger than 0.9, which shows a high performance of the ps2code image classifier of \emph{psc2code}. 
For example, for the playlist P1, our model can identify most of valid and invalid frames. 
However, the accuracy of our approach on some cases is not very good. For example, for the playlist P11, the accuracy is only 0.38. In this playlist, the NetBeans IDE is used and the image classifier fails to recognize many frames with code completion popups as noisy-code frames. This may be because of the limited number of training data that we have for NetBeans IDE.

In our frame classification task, we care more about the performance on valid code frames.
On the other hand, misclassifying too many valid code frames as invalid may result in information loss in the OCRed code for the video.
In terms of recall of valid code frames, the overall recall is 0.85.
For 14 out the 23 plalists, the recall is greater than 0.9 for 14 playlists. In two cases (e.g., P7 and P17), the recall is equal to 1. 
However, for the playlist P11, the recall is very low, i.e., 0.13, which might again be caused by the NetBeans IDE used in its programming videos and the limited number of training data which uses the NetBeans IDE. 
For the remaining playlists, we find that for the most of misclassified valid frames, there are ``nearby'' valid code frames (i.e., the timestamps of the frames are close to those of the misclassified frames) that have the same or similar code content. 
We also find that anthoer reason for misclassified frames is the intermediate code. For example, one misclassified code frame contains a line of code {\tt System.out.println("")}, which is the intermediate code of {\tt System.out.println("Copy of list :")} in a later frame. 
Overall, the information loss resulting from misclassifying valid code frames as invalid is acceptable.

On the other hand, misclassifying too many invalid frames as valid may result in much noise in the OCRed code.
In terms of precision on valid frames, the overall precision is 0.91 with 16 out of 23 playlists have precision scores above 0.9.
Playlist 17 had the lowest precision with 0.66. We find that there are many frames in which the console window overlaps with the code editor, thus our image classifier misclassifies them because this type of invalid frames does not appear in the training data.
Overall, the negative impact of misclassifying invalid frames as valid on the subsequent processing steps is minor, but more training data can further improve the model's precision on valid code frames.

\subsubsection{Effectiveness of Locating Code Regions in Code Frames (RQ3)}\label{sec:rq3}


\noindent \textbf{Approach.}

We use the approach of Alahmadi et al.~\cite{alahmadi2018accurately} as our baseline. Their approach leverages the You Only Look Once (YOLO) neural network~\cite{redmon2016you}, which can predict the location of code fragments in programming video tutorials directly. Given a frame, the baseline approach can not only predict whether it is valid or not, but also can identify the location of code fragments for valid frames. We use the labeled frames from Section~\ref{sec:training} to train a model for the baseline and apply the trained model on these 4,828 frames.

Additionally, we use the Intersection over Union (IoU) metric~\cite{zitnick2014edge}, which is also used in the Alahmadi et al.'s work, to measure the accuracy of a predicted code bounding box for a frame. IoU divides the area of overlap between the bounding boxes of the prediction and ground truth by the area of the union of the bounding boxes of the prediction and ground truth. 
Given a frame, we use the same IoU computation as the Alahmadi et al.'s work, which is as follows:
\begin{itemize}
    \item If a model predicts it as an invalid frame and it is correct, then $IoU=1$.
    \item If the prediction result of a model is incorrect, then $IoU=0$.
    \item If a model predicts it as a valid frame and it is correct, then $IoU=\frac{A_{gt}\cap A_{pred}}{A_{gt}\cup A_{pred}}$, where $A_{gt}$ and $A_{pred}$ are the area of the ground truth and the predicted bounding boxes respectively.
\end{itemize}
We compute the average of IoU for each playlist and an overall IoU on all the frames.

To generate the ground truth of the bounding boxes that contain code, we provide a web application that show the frames with the predicted code area and allow the annotators to adjust the bounding box. When two annotators were labeling a valid frame, they were also required to adjust the bounding box of its predicted code area. 
When the IoU of the bounding boxes labeled by two annotators for a frame is larger than 95\%, which indicates most of the two bounding boxes overlap, we think that the two annotators reach an agreement. 
Since the ground truth of the bounding boxes that contain code is usually the code editor in the IDE (see Figure~\ref{tbl:example}(a) and (d)), the two annotators reach an agreement easily for most of valid frames. 
A small number of disagreement cases were caused by the margin of the code editor. Most of the disagreement cases are due to the scroll bars in the code editor window. After a discussion, the two annotators include the scroll bars in the annotated code area for all the cases. Compared with the whole code editor windows, the areas of scroll bars are small. Thus, our results and findings would not be affected by much.

\noindent \textbf{Results.}
Table~\ref{tbl:comparison} presents the comparison results between \emph{psc2code} and the baseline approach on accuracy, F1-score on valid and invalid frames, and IoU.
In terms of accuracy and F1-score, the image classifier of \emph{psc2code} achieves much better performance than the baseline for most of the playlist except for the playlist P3, P7, and P18. But for the playlist P3, P7 and P18, the differences are very small. 
In terms of IoU, all values are larger than 0.85 except the playlist P1. For the playlist P1, the value of IoU is 0.78, which is also considered as a successful prediction in the study of Alahmadi et al.~\cite{alahmadi2018accurately}. 
Comparing to our approach, the IoUs achieved by the baseline are much smaller. We apply Wilcoxon signed-rank test~\cite{wilcoxon1945individual} and find that the differences are statistically significant at the confidence level of 99\%.


The low IoUs of the baseline is caused by its incorrect predicted results. Moreover, even if we ignore the incorrect cases, the overall IoU is still 0.76, which is smaller than that of our approach. To get more insight, we look into the valid code frames with bounding boxes identified by the baseline. We found that the bounding boxes identified by the baseline usually did not cover the whole area of the code editor and missed some parts of the code area, which result in information loss. Figure~\ref{fig:baseline} shows an example identified by the baseline. We can see that the bounding boxes of the two frames identified by the baseline miss some code content.

\begin{figure}
    \includegraphics[width=0.7\textwidth]{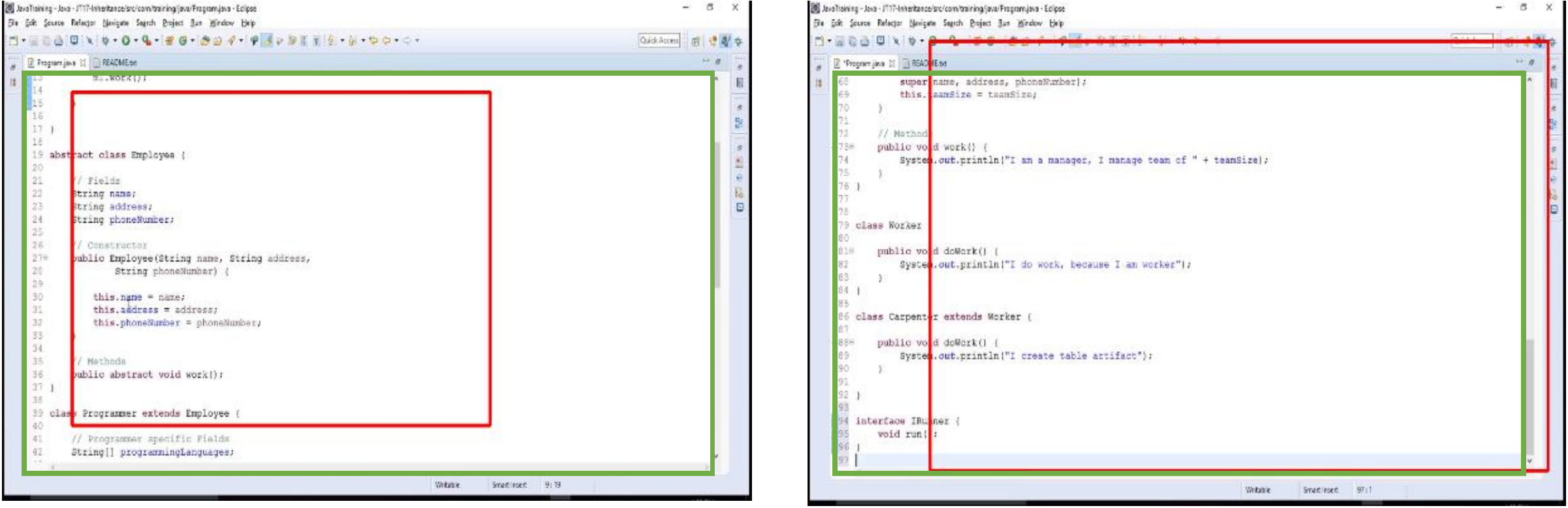}
    \caption{Two example frames with bounding boxes identified by the baseline (red color) from the playlist P3; the ground truth of the bounding boxes is in green color.}\label{fig:baseline}
\end{figure}

Although YOLO is a powerful object detection model for generic object detection, but we think there are some limitations of object detection based methods for our code extraction task. 
First, the boundary features of strokes, characters, words and text lines can confuse the YOLO model for accurately detecting code window boundaries (see Fig. 8 for example).
Second, YOLO uses a set of anchor boxes (with pre-defined sizes and aspect ratios) to locate the potential object regions. This works fine for natural objects (e.g., person, dog, car), but it is fundamentally limited for locating code content which can have arbitrary sizes and aspect ratios. Third, YOLO uses CNN to extract image features. Due to the CNN's spatial downsampling nature, it can only produce an approximate bounding box around the object (with some background and even parts of other objects in the box). This is fine for detecting natural objects as long as the boxes cover a large region of the object. However, such approximate bounding boxes will not satisfy the much higher accuracy requirement of locating code regions, as we do not want to miss any actual code fragments and do not want to include any non-code regions

\subsubsection{Improvement of the Quality of the OCRed Source Code (RQ4)}

\medskip \noindent \textbf{Approach.}
In this RQ, we use the same 46 programming videos examined in the RQ2.
To evaluate the ability of our approach to correct OCR errors in the OCRed source code, we first get a list of distinct words from the OCRed source code for each video.
Then, we determine the correctness of the words based on the statistical language model learned from the source-code corpus and count the number of correct and incorrect words.
For the incorrect words, we correct them using the methods proposed in Section~\ref{sec:step4}, and count the number of incorrect words that are corrected by our approach.
Among the corrected words, some words may be not truly corrected.
For example, given several similar variables in the code such as {\tt obj1, obj2, obj3}, these variables might be OCRed as {\tt obji} or {\tt objl} due to the overlapping cursor or other noise. 
Thus, our approach may choose a wrong word from these similar candidate words based on cross-frame information.
Thus, we manually check whether a word with OCR errors is truly corrected by comparing it with the content of the corresponding frame.
We calculate two correction accuracies: the proportion of the truly corrected words in all incorrect words (Accuracy1) and the proportion of the truly corrected words in the words corrected by our approach (Accuracy2)

\medskip \noindent \textbf{Results.}
Table~\ref{tbl:rq2} presents the analysis results. 
In this table, the column \#CorrectOCR and \#ErrorOCR list the number of distinct correct and incorrect words identified by the statistical language model, respectively.
The column \#Corrected is the number of incorrect words corrected by our approach and the column \#TrueCorrected is the number of incorrect words that are truly corrected by our approach.
The last two columns list Accuracy1 and Accuracy2 respectively.

As shown in Table~\ref{tbl:rq2}, there are many words with OCR errors in the OCRed source code. 
Overall, the ratio of incorrect words and correct words is about 1:3.
Among all incorrect words, our approach makes corrections for half of them.
For the incorrect words that our approach does not attempt to correct, many of them are partial words while the developer is still typing the whole word, and some are meaningless words resulting from the noise in the frame.
Among the incorrect words that are corrected by our approach, most of them (88\%) are truly corrected.
Many words are falsely corrected when there are multiple similar words in the code such as similar variables {\tt obj1, obj2, obj3}. 
Overall, our approach can truly correct about half of incorrect words (46\%), which can significantly improve the quality of the OCRed source code by reducing the ratio of incorrect words and correct words from 1:3 to 1:6.


\begin{table*}[]
    \centering
    \caption{The statistics of the OCRed source code corrected by \emph{psc2code}  }\label{tbl:rq2}
    \begin{tabular}{@{}lrrrrrr@{}}
    \toprule
    Playlist & \#CorrectOCR & \#ErrorOCR & \#Corrected & \#TrueCorrected & Accuracy1 & Accuracy2 \\ \midrule
    P1       & 166       & 34         & 25          & 24              & 0.71   & 0.96   \\
    P2       & 121       & 29         & 13          & 8               & 0.28   & 0.62   \\
    P3       & 175       & 30         & 9           & 7               & 0.23   & 0.78   \\
    P4       & 55        & 15         & 4           & 4               & 0.27   & 1.00   \\
    P5       & 284       & 176        & 132         & 92              & 0.52   & 0.70   \\
    P6       & 95        & 16         & 15          & 15              & 0.94   & 1.00   \\
    P7       & 85        & 59         & 40          & 37              & 0.63   & 0.93   \\
    P8       & 237       & 81         & 30          & 21              & 0.26   & 0.70   \\
    P9       & 181       & 143        & 91          & 85              & 0.59   & 0.93   \\
    P10      & 75        & 14         & 9           & 7               & 0.50   & 0.78   \\
    P11      & 149       & 102        & 31          & 31              & 0.30   & 1.00   \\
    P12      & 507       & 184        & 86          & 82              & 0.45   & 0.95   \\
    P13      & 195       & 28         & 11          & 11              & 0.39   & 1.00   \\
    P14      & 178       & 53         & 33          & 33              & 0.62   & 1.00   \\
    P15      & 329       & 119        & 61          & 52              & 0.44   & 0.85   \\
    P16      & 39        & 3          & 1           & 1               & 0.33   & 1.00   \\
    P17      & 159       & 58         & 14          & 13              & 0.22   & 0.93   \\
    P18      & 179       & 45         & 24          & 22              & 0.49   & 0.92   \\
    P19      & 136       & 17         & 8           & 8               & 0.47   & 1.00   \\
    P20      & 99        & 20         & 7           & 7               & 0.35   & 1.00   \\
    P21      & 59        & 9          & 4           & 4               & 0.44   & 1.00   \\
    P22      & 162       & 79         & 42          & 40              & 0.51   & 0.95   \\
    P23      & 158       & 43         & 25          & 24              & 0.56   & 0.96   \\  \midrule
    All      & 3,823      & 1,357       & 715         & 628             & 0.46   & 0.88   \\ \bottomrule
    \end{tabular}
    \end{table*}


\subsubsection{Efficiency of  Our Approach(RQ5)}

\medskip \noindent \textbf{Approach.}
In this RQ, we apply our approach on the 46 programming videos used in RQ1. As shown in Section~\ref{sec:approach}, our approach has four steps to process a programming video: 1) Reducing Non-Informative Frames, 2) Removing Non-Code and Noisy-Code Frames, 3) Distinguishing Code versus Non-Code Regions, 4) Correcting Errors in OCRed Source Code. So, given a programming video, we compute the time used by each step in our approach. We run our approach on a machine with Intel Core i7 CPU, 64GB memory, and one NVidia 1080Ti GPUs with 16 GBs of memory.

\medskip \noindent \textbf{Results.}
Table~\ref{tbl:efficiency} presents the statistics of run time of each step of our approach on these 46 programming videos. \emph{psc2code} takes 47.64 seconds to complete processing a programming video on average. The first step, i.e., reducing non-informative frames, takes the longest time. This is because it need to process all frames in a programming videos. We find that the process time has a positive relationship with the duration of the programming videos. For example, the duration of a video in the playlist P5 is about 1 hour and 41 minutes; thus \emph{psc2code} takes about 493 seconds to complete the process including 339 seconds in the first step. For the other three steps, it is fast for \emph{psc2code} to complete the process. On average, each step only needs less than 10 seconds. 
In sum, we believe that our approach can efficiently process a large number of programming videos.

\begin{table}[]
    \caption{The statistics of run time (seconds) of each step of \emph{psc2code} on 46 programming videos.}\label{tbl:efficiency}
    \begin{tabular}{@{}lrrrrr@{}}
    \toprule
           & Step 1 & Step 2 & Step 3 & Step 4 & All \\ \midrule
    Mean   & 32.04  & 5.93   & 7.19   & 2.48   & 47.64 \\
    Std.   & 50.67  & 5.74   & 16.57  & 4.51   & 73.16 \\
    Median & 19.53  & 4.64   & 3.17   & 0.90   & 27.21 \\ \bottomrule
    \end{tabular}
\end{table}

\section{Applications}\label{sec:app}

In this section, we describe two downstream applications built on the source code extracted from programming screencasts by \emph{psc2code}: programming video search and enhancing programming video interaction.

\subsection{Programming Video Search}

As a programming screencast is a sequence of screen images, it is difficult to search programming videos by their image content.
A primary goal of extracting source code from programming screencasts is to enable effective video search based on the extracted source code.

\subsubsection{Programming Video Search Engine}
In this study, we build a programming video search engine based on the source code of the 1142 programming videos extracted by \emph{psc2code}.
For each programming video in our dataset, we aggregate the source code of all its valid code frames as a document. The source-code documents of all 1142 videos constitutes a source-code corpus.
We refer to this source-code corpus as the denoisied corpus as opposed to the noisy source-code corpus produced by the baseline method described below.
We then compute a TF-IDF (Term Frequency and Inverse Document Frequency) vector for each video in which each vector element is the TF-IDF score of a token in the source-code document extracted from the video.
Given a query, the search engine finds relevant programming videos by matching the keywords in the query with the tokens in the source code of programming videos.
The returned videos are sorted by the descending order of the sum of the the TF-IDF scores of the matched source-code tokens in the videos.
For each returned programming video, the search engine also returns the valid code frames that contain any keyword(s) in the query.


\subsubsection{Baseline Methods for Source-Code Extraction}

To study the impact of the denoising features of \emph{psc2code} on the downstream programming video search, we build two baseline methods to extract source code from programming videos:
\begin{itemize}
    \item 
    The first baseline method (which we refer to as \emph{baseline1}) does not use the denoising features of \emph{psc2code}. To implement this baseline, we follow the same strategy as  \emph{psc2code} in the removal of  redundant frames and detection of code regions. However,  we do not remove the noisy-code frames and non-code frames. For this baseline, we use Google Vision API to extract the source code and check if the detected subimages have code or not by identifying whether there exist Java keywords in the extracted text; this is the same method used by Kandarp and Guon~\cite{khandwala2018codemotion}. In this way,  the baseline removes non-code frames. Different from \emph{psc2code}, \emph{baseline1} does not remove noisy-code frames, nor does it fix the errors in the OCRed source code. baseline1 uses the two heuristics of Kandarp and Guon~\cite{khandwala2018codemotion} to remove line numbers and Unicode errors.
    
    \item We use the full-fledged CodeMotion~\cite{khandwala2018codemotion} as the second baseline method, which is referred to as \emph{baseline2}. We obtain the source code of CodeMotion from CodeMotion's first author's Github repository.
\end{itemize}

For each baseline method, we obtain a noisy source-code corpus for the 1142 programming videos in our dataset which we use for the comparison of the search results quality against the denoised source-code corpus created by \emph{psc2code}.

\subsubsection{Search Queries}
The majority of videos in our video corpus introduce basic concepts in Java programming. In a typical playlist of a Java programming tutorial (e.g., the playlist P1), the creator usually first introduces syntax of Java and the concept of object-oriented programming (e.g., the java.lang.Object class), then he/she writes code to introduce some common used classes, such as String, Collections, File; Finally, some advanced knowledge (e.g., multi-thread and GUI programming) may be introduced. 

Based on the topics discussed in the programming video tutorials in our dataset, we select the common used classes and APIs in our video corpus to design 20 queries (see Table~\ref{tbl:rq31}), which cover three categories of programming knowledge, including basic Java APIs (e.g., string operations, file reading/writing, Java collection usages), GUI programming (e.g., GUI components, events and listeners), and multi-threading programming (e.g., thread wait/sleep).
For these queries, there are reasonable numbers of programming videos in our dataset.
This allows us to investigate the impact of the denoised source code by \emph{psc2code} on the quality of the search results, compared with the noisy source code extracted by the baseline.

\subsubsection{Evaluation Metrics}

In this study, only if all the keywords in a query can be found in at least one valid code frame of a video, the returned video is considered as truly relevant to the query.
To verify whether the returned videos are truly relevant to a query (i.e., the video truly demonstrates the usage of the class and API referred to in the query), the first two authors manually and carefully check the top-20 videos in the search results for the 20 search queries.

To compare the search results quality on the denoised source-code corpus by \emph{psc2code} and the noisy source-code corpus by the baseline, we use several well-known evaluation metrics: precision@k, MAP@k (Mean Average Precision~\cite{baeza1999modern}), and MRR@k (Mean Reciprocal Rank~\cite{baeza1999modern}), which are commonly used in past studies involving building recommendation systems~\cite{rao2011retrieval, tamrawi2011fuzzy, zhou2012should, xia2015automatic, xia2017effective}.

For each query $Q_i$, let $V_i$ be the number of videos that are truly relevant to the query in the top-k videos returned by a video search engine over a source-code corpus extracted from a set of programming videos.
The precision@k is the ratio of $V_i$ over $k$, i.e. $\frac{V_i}{k}$.

MAP considers the order in which the returned videos are presented in the search results.
For a single query, its \textit{average precision (AP)} is defined as the mean of the precision values obtained for different sets of top-j videos that are returned before each relevant video in the search results, which is computed as:
$$AP=\frac{\sum_{j=1}^{n}{P(j)\times Rel(j)}}{\text{the number of relevant videos}}$$
where $n$ is the number of returned videos, $Rel(j)=1$ indicates that the video at position $j$ is relevant, otherwise $Rel(j)=0$, and $P(j)$ is the precision at the given cut-off position $j$.
Then, for the $m$ queries, the MAP is the mean of $AP$, i.e., $\frac{\sum_{i=1}^{m}{AP_i}}{m}$.

Different from MAP that considers all relevant videos in the search results, MRR considers only the first relevant video. Given a query, its reciprocal rank is the multiplicative inverse of the rank of the first relevant video in a ranked list of videos. For the $m$ queries, MRR is the average of their reciprocal ranks, which is computed as:
$$MRR=\frac{1}{m}\sum{\frac{1}{rank(q)}}$$
where $rank(q)$ refers to the position of the first relevant video in the ranked list returned by the search engine.

\begin{table*}[]
    \centering
    \caption{The performance of the search engines built on \emph{psc2code} and the two baselines.}\label{tbl:rq31} 
    \resizebox{\textwidth}{!}{%
    \begin{tabular}{@{}l|rrrrrrrrr@{}}
    \toprule
    \multirow{2}{*}{Query} & \multicolumn{3}{c|}{precision@5}                                          & \multicolumn{3}{c|}{precision@10}                     & \multicolumn{3}{c}{precision@20} \\
                           & \multicolumn{1}{r}{psc2code} & baseline1 & \multicolumn{1}{r|}{baseline2} & psc2code & baseline1 & \multicolumn{1}{r|}{baseline2} & psc2code & baseline1 & baseline2 \\ \midrule
    ArrayList isEmpty      & 1.00                         & 0.00      & \multicolumn{1}{r|}{0.60}      & 1.00     & 0.10      & \multicolumn{1}{r|}{0.30}      & 0.50     & 0.15      & 0.15      \\
    Date getTime           & 0.60                         & 0.60      & \multicolumn{1}{r|}{0.60}      & 0.30     & 0.30      & \multicolumn{1}{r|}{0.30}      & 0.15     & 0.15      & 0.15      \\
    event getSource        & 1.00                         & 1.00      & \multicolumn{1}{r|}{0.80}      & 1.00     & 1.00      & \multicolumn{1}{r|}{0.90}      & 1.00     & 0.90      & 0.90      \\
    File write             & 1.00                         & 0.60      & \multicolumn{1}{r|}{0.60}      & 0.80     & 0.30      & \multicolumn{1}{r|}{0.40}      & 0.40     & 0.40      & 0.30      \\
    HashMap iterator       & 1.00                         & 0.60      & \multicolumn{1}{r|}{0.80}      & 0.70     & 0.50      & \multicolumn{1}{r|}{0.60}      & 0.35     & 0.35      & 0.35      \\
    imageIO read file      & 1.00                         & 0.60      & \multicolumn{1}{r|}{0.20}      & 1.00     & 0.50      & \multicolumn{1}{r|}{0.10}      & 0.55     & 0.55      & 0.05      \\
    Iterator forEach       & 1.00                         & 0.60      & \multicolumn{1}{r|}{0.60}      & 0.80     & 0.40      & \multicolumn{1}{r|}{0.50}      & 0.40     & 0.35      & 0.40      \\
    Iterator remove        & 1.00                         & 0.60      & \multicolumn{1}{r|}{0.80}      & 1.00     & 0.60      & \multicolumn{1}{r|}{0.90}      & 0.90     & 0.55      & 0.80      \\
    JButton keyListener    & 1.00                         & 0.50      & \multicolumn{1}{r|}{0.40}      & 0.50     & 0.40      & \multicolumn{1}{r|}{0.20}      & 0.25     & 0.25      & 0.10      \\
    JFrame setLayout       & 1.00                         & 0.80      & \multicolumn{1}{r|}{1.00}      & 1.00     & 0.80      & \multicolumn{1}{r|}{0.80}      & 1.00     & 0.90      & 0.90      \\
    JFrame setSize         & 1.00                         & 1.00      & \multicolumn{1}{r|}{1.00}      & 1.00     & 1.00      & \multicolumn{1}{r|}{1.00}      & 1.00     & 1.00      & 1.00      \\
    List indexOf           & 0.80                         & 0.20      & \multicolumn{1}{r|}{0.60}      & 0.40     & 0.20      & \multicolumn{1}{r|}{0.30}      & 0.20     & 0.10      & 0.15      \\
    List sort              & 1.00                         & 0.40      & \multicolumn{1}{r|}{0.80}      & 1.00     & 0.60      & \multicolumn{1}{r|}{0.80}      & 1.00     & 0.55      & 0.65      \\
    Object getClass        & 1.00                         & 0.20      & \multicolumn{1}{r|}{0.80}      & 1.00     & 0.40      & \multicolumn{1}{r|}{0.50}      & 1.00     & 0.30      & 0.30      \\
    Object NullException   & 1.00                         & 0.60      & \multicolumn{1}{r|}{0.80}      & 1.00     & 0.70      & \multicolumn{1}{r|}{0.90}      & 0.95     & 0.80      & 0.90      \\
    String concat          & 1.00                         & 0.40      & \multicolumn{1}{r|}{1.00}      & 1.00     & 0.30      & \multicolumn{1}{r|}{0.60}      & 0.55     & 0.30      & 0.35      \\
    String format          & 1.00                         & 0.40      & \multicolumn{1}{r|}{0.80}      & 1.00     & 0.40      & \multicolumn{1}{r|}{0.70}      & 1.00     & 0.45      & 0.55      \\
    StringBuffer insert    & 0.40                         & 0.20      & \multicolumn{1}{r|}{0.20}      & 0.20     & 0.10      & \multicolumn{1}{r|}{0.10}      & 0.10     & 0.05      & 0.10      \\
    thread wait            & 0.80                         & 0.40      & \multicolumn{1}{r|}{0.40}      & 0.40     & 0.30      & \multicolumn{1}{r|}{0.30}      & 0.20     & 0.20      & 0.20      \\
    thread sleep           & 1.00                         & 1.00      & \multicolumn{1}{r|}{1.00}      & 1.00     & 1.00      & \multicolumn{1}{r|}{0.90}      & 1.00     & 0.95      & 0.95      \\ \midrule
    Average                & 0.93                         & 0.53      &     0.69                           & 0.81     & 0.50      &   0.56                             & 0.63     & 0.46      &   0.46        \\ \bottomrule
    \end{tabular}
    }
\end{table*}

\begin{table}[]
\centering
\caption{The MAP@k and MRR@k of \emph{psc2code} and the baseline}\label{tbl:rq32}
\begin{tabular}{@{}lrrr@{}}
\toprule
         & MAP@5 & MAP@20 & MAP@20 \\ \midrule
psc2code  & 0.998 & 0.998  & 0.996  \\
baseline1 & 0.700 & 0.654  & 0.620  \\ 
baseline2 & 0.813 & 0.786  & 0.763  \\ \midrule \midrule
         & MRR@5 & MRR@10 & MRR@20 \\ \midrule
psc2code  & 1.000 & 1.000  & 1.000  \\
baseline1 & 0.789 & 0.788  & 0.788  \\
baseline2 & 0.825 & 0.825  & 0.825 \\ \bottomrule
\end{tabular}
\end{table}

\subsubsection{Results}

We compute the precision@k, MAP@k and MRR@k metrics for the top-5, top-10 and top-20 search results (i.e., k=5, 10 and 20).
Table~\ref{tbl:rq31} presents the results of precision@k for each query and Table~\ref{tbl:rq32} presents the results of MAP@k and MRR@k for the 20 queries.
We can observe that the quality of the search results for the denoised source-code corpus extracted by our \emph{psc2code} is much higher compared to the search results we obtained on the noisy source-code corpus extracted from the two baselines.

The average precision@5, precision@10 and precision@20 of our approach over the 20 queries are 0.93, 0.81, and 0.63, respectively. The average precision@5, precision@10 and precision@20 of \emph{baseline1} are only 0.53, 0.50, and 0.46, respectively, while the average precision@5, precision@10 and precision@20 of \emph{baseline2} are only 0.69, 0.56, and 0.46, respectively.
The precision@5 of our approach is 1 for 16 out of the 20 queries.
That is, for these 16 videos, the top-5 videos returned based on our denoised source-code corpus are all relevant to the search query for these 16 queries.
For the other four queries (``Data getTime'', ``StringBuffer insert'', ``List indexOf'' and ``thread wait''), some of the top-5 returned videos are not relevant to the query.
The reason is that the number of programming videos in our dataset is limited (i.e., 1142).
As such, usually only a small number of videos contain the searched APIs.
For example, we check the source code of all the videos in our dataset and find that only two programming videos mention the API {\tt StringBuffer.insert}.
In fact, these two videos are returned in the top-5 results for the query ``StringBuffer insert''. But as there are only these two truly relevant videos in the whole dataset, the precision@5 is only 0.4 for this query.

We note that both our approach and the baselines have high precision in the search results for some queries, e.g., ``event getSource'' and ``thread sleep''.
This is because APIs relevant to these queries are always used together in the source code of the programming screencasts.
For such cases, searching programming videos based on the source code extracted by the baseline may achieve acceptable precision.
However, the precision of the baseline is generally not satisfactory for most of the queries.
The precision@5 based on the noisy source-code corpus by the baseline is 1 only for three queries.
For the query ``ArrayList isEmpty'', there are no relevant videos returned in the top-5 search results (i.e., precision@5=0).
This is because when a developer uses the class ArrayList, the API ``isEmpty'' frequently appears in the completion suggestion popups but is not actually used in the real code.
This noise in the extracted code subsequently leads to less accurate programming video search.

All the MAPs and MRRs of our approach are equal to or close to 1, which indicates that the search engine can rank the truly relevant videos in the dataset at the top of the search results.
For some queries, when $k$ increases, the top-k precision of our approach decreases.
This is because our video dataset may only have a small number of videos relevant to a query and those ranked lower in the search results are mostly irrelevant to the query.
But when the dataset has a large number of relevant programming videos for a query (e.g., ``list sort'', ``thread sleep''), the precision@10 or even the precision@20 can still be 1, which means that a large number of relevant programming videos (if any) can be found and ranked in the top search results.
In contrast, the MAPs and MRRs of the baselines are much lower than those of our approach.
For some queries (e.g., ``Object NullException''), it is interesting to note that the top-k precision of the baseline increases when $k$ increases.
This is because the truly relevant videos may be ranked below the irrelevant videos in the search results due to the noise source code extracted by the baseline.

\textbf{Summary:} the search engine based on the denoised source-code corpus extracted by our approach can recommend programming videos more accurately than the two baseline methods. 
In the future, we plan to add more information of the programming videos such as their textual description, audio, etc to improve the search results, which is similar to CodeTube~\cite{ponzanelli2016too} and VT-Revolution~\cite{bao2018vt}.

\begin{figure*}[t]
    \centering
    \includegraphics[width=\textwidth]{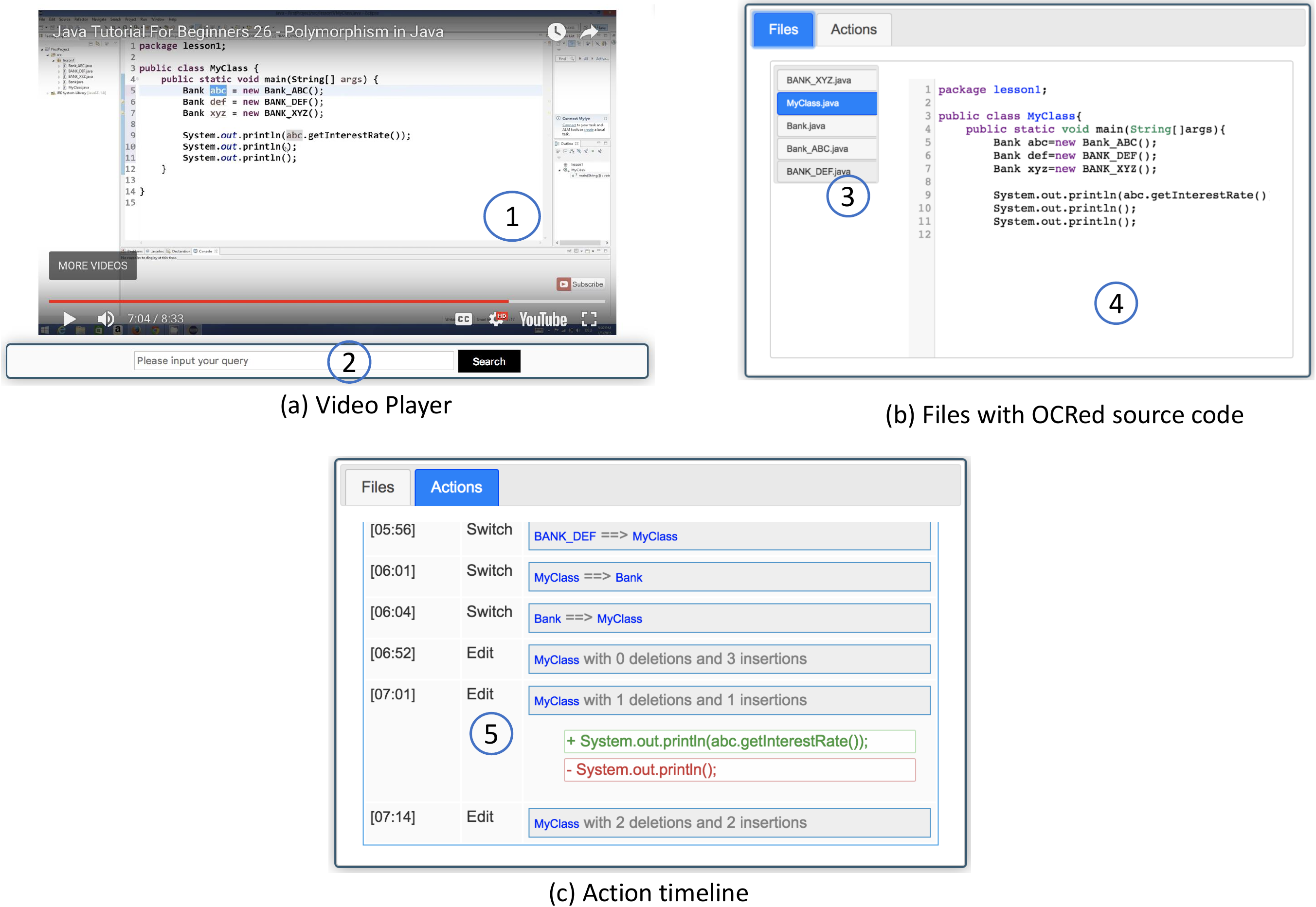}
    \caption{The screenshots of a YouTube video enhanced by \emph{psc2code}. (1) Video Player. (2) Search/navigate by code content. (3) Identified Files. (4) File content is synchronous with the video playing. (5) Action timeline.}\label{fig:enhanced}
\end{figure*}

\subsection{Enhancing Programming Video Interaction}
Due to the image streaming nature of programming videos, it is difficult to navigate and explore the content of programming videos.
In our previous work, we proposed a tool named VTRevolution~\cite{bao2018vt}, which enhances the navigation and exploration of programming video tutorials by providing programming-specific workflow history and timeline-based content browsing.
But VTRevolution needs to collect the developer's human-computer interaction data through system level instrumentation when creating a programming video tutorial.
Thus, it is impossible to directly apply the interaction-enhancing features of VTRevolution to the large amount of existing programming screencasts on the Internet.
Based on the source code extracted by \emph{psc2code}, we can use a similar interaction design of VTRevolution~\cite{bao2018vt} to make existing programming screencasts more interactive and ease the navigation and exploration of programming video content.

\subsubsection{Prototype Design}

We developed a web application prototype that leverages the source code of existing programming video tutorials on YouTube extracted by \emph{psc2code} to enhance the navigation and exploration of programming videos.
Figure~\ref{fig:enhanced} shows the screenshots of this prototype tool, which supports the following interaction-enhancing features:

\vspace{0.1cm} \noindent \textbf{Search/navigate the video by code content.}
This feature allows video viewers to find the code elements that they are interested in based on the OCRed source code. Tutorial watchers enter a query in the search box (annotation \textcircled{2} in Fig.~\ref{fig:enhanced} (a)). The prototype currently performs a simple substring match between the entered keywords and the OCRed source code of the valid code frames in the video. It returns a list of time stamps of the matched frames in a chronological order. Tutorial watchers can double-click a time stamp in the results list to navigate the tutorial video to the frame of the double-clicked time stamp.

\vspace{0.1cm} \noindent \textbf{View file content.}
This feature provides a file content view (annotation \textcircled{4} in Fig.~\ref{fig:enhanced} (b)) that allows tutorial watchers to view all contents that the tutorial author already enters to a file till the current time of video playing during a programming tutorial.
To detect the different files in a video tutorial, we use the density-based clustering algorithm DBSCAN~\cite{ester1996density} to cluster the frames based on difference of their lines of code (LOC). Given two frames, we compute the Longest Common Lines between their LOC, which is similar to Longest Common String (LCS). Then, we normalized the number of longest common lines by dividing the number of LOC of the frame with longer LOC as the dissimilarity between two frames. Additionally, we regard each cluster as a file opened in the video tutorial and identify its name using the class name if the file contains a Java class.
As shown in annotation \textcircled{3} in Fig.~\ref{fig:enhanced} (b), five Java files are identified in the video till the current time of video playing.
We find that the clustering algorithm can identify the files in the three videos used in the user study (see Section~\ref{sec:userstudy}) correctly. We identified 3, 5, and 2 Java files for the three videos, respectively.
In the file content view, the focused file (i.e., the file currently visible in the video) and its content is synchronized with the video playing. As changes are made to the focused file in the video, the code content of the focused file will be updated automatically in the file content view. Although only a part of the focused file is visible in the current video frame, tutorial watchers can view the whole already-created file content of the focused file in the file content view, and also switch to non-focused files and view their contents, without the need to navigate the video to the frame where that content is visible. 
Unfortunately, we cannot guarantee that the whole content of the source code is complete or correct because there are many complicated cases (e.g., some contents are never shown in the video), which are difficult for our prototype to handle.

\vspace{0.1cm} \noindent \textbf{Action timeline.}
This feature allows users to view when the tutorial author does what to which file. To detect actions in the programming video, we compare the OCRed source code between the adjacent valid code frames processed by \emph{psc2code}. If the adjacent frames belong to the same file and the extracted code content is different, the action is denote as an \emph{edit} with a summary of the number of inserted and deleted code lines.
The code difference can be viewed by clicking to expand an edit action in the prototype (see Figure~\ref{fig:enhanced} (c)).
The prototype uses $+$ sign and green color to indicate code being inserted, and $-$ sign and red color to indicate code being deleted.
If the adjacent frames belong to the two different files, the action is denoted as a \emph{switch} from one file to another.
Tutorial watchers can click the time stamp of an action to navigate the programming video to that time when the action occurs.

\subsubsection{User Study Design}\label{sec:userstudy}
We conducted a user study to evaluate the applicability of our prototype tool for enhancing the developers' interaction with programming videos.

\begin{table*}[]
    \centering
    \caption{The selected programming videos and their corresponding questionnaire used in the user study.}\label{tbl:app2}
    \resizebox{1.0\textwidth}{!}{%
    \begin{tabular}{@{}lllll@{}}
    \toprule
    Index               & Video Title                                                                                                                          & Duration               & Questions                                                                                                         & Category  \\ \midrule
    \multirow{5}{*}{V1} & \multirow{5}{*}{\begin{tabular}[c]{@{}l@{}}3 - Canvas - New Beginner 2D\\ Game Programming (\href{https://www.youtube.com/watch?v=ck39jt04Qpk}{Video Link})\end{tabular}}          & \multirow{5}{*}{06:13} & Q1. How many class files are opened and viewed in this programming video?                                         & Content   \\ \cline{4-5}
                        &                                                                                                                                 &                        & Q2. Which classes have the main entry?                                                                            & Content   \\ \cline{4-5}
                        &                                                                                                                                 &                        & Q3. Which classes are newly created in the video?                                                                 & Workflow  \\ \cline{4-5}
                        &                                                                                                                                 &                        & Q4. How to set size for a JFrame instance in the video?                                                           & API Usage \\ \cline{4-5}
                        &                                                                                                                                 &                        & Q5. How to set size for a Canvas instance in the video?                                                           & API Usage \\ \midrule
    \multirow{5}{*}{V2} & \multirow{5}{*}{\begin{tabular}[c]{@{}l@{}}Java Tutorial For Beginners 26\\ - Polymorphism in Java (\href{https://www.youtube.com/watch?v=GnLtvmeGAWA}{Video Link})\end{tabular}}  & \multirow{5}{*}{08:33} & Q1. How many class files are opened and viewed in this programming video?                                         & Content   \\  \cline{4-5}
                        &                                                                                                                                 &                        & Q2. Which classes are the sub class of class Bank?                                                                & Content   \\  \cline{4-5}
                        &                                                                                                                                 &                        & Q3. What is return value of the function getInterestRate in the class Bank\_ABC?                                   & Output    \\  \cline{4-5}
                        &                                                                                                                                 &                        & Q4. What is return value of the function getInterestRate in the class Bank\_GHI?                                   & Output    \\  \cline{4-5}
                        &                                                                                                                                 &                        & Q5. What is return value of the function getInterestRate in the class Bank\_XYZ?                                   & Output    \\ \midrule
    \multirow{5}{*}{V3} & \multirow{5}{*}{\begin{tabular}[c]{@{}l@{}}Java Tutorial for Beginners - 31\\ - Getters and Setters (\href{https://www.youtube.com/watch?v=OF3vBYWikYs}{Video Link})\end{tabular}} & \multirow{5}{*}{12:00} & Q1. How many class files are opened and viewed in this programming video?                                         & Content   \\  \cline{4-5}
    &                                                                                                                                 &                        & Q2. \begin{tabular}[c]{@{}l@{}}What can be the value of the field Orc.height by calling getHeight(9)\\when the author creates a function getHeight(int) initially\end{tabular}  & Workflow  \\  \cline{4-5}
                        &                                                                                                                                 &                        & Q3. \begin{tabular}[c]{@{}l@{}}The author revises the initial getHeight(int) afterwards,\\what changes are made to this function?\end{tabular} & Workflow  \\  \cline{4-5}
                        &                                                                                                                                 &                        & Q4. \begin{tabular}[c]{@{}l@{}}What can be the value of the field Orc.height by calling setHeight(9)\\using the first version of setHeight(int)?\end{tabular} & Workflow  \\  \cline{4-5}
                        &                                                                                                                                 &                        & Q5. \begin{tabular}[c]{@{}l@{}}When calling setHeight(9) by two different versions of setHeight(int),\\will be value of Ocr.Height be set as the same value?\end{tabular} & Workflow  \\ \bottomrule
    \end{tabular}
    }
\end{table*}

\vspace{0.1cm} \noindent \textbf{Video Selection.}
We selected three programming videos from our video dataset in Table~\ref{tbl:rq1}.
As summarized in Table~\ref{tbl:app2}, the duration of these three videos is representative of YouTube programming videos (from 6 minutes to 12 minutes).
These three videos cover different programming topics:
the first video presents the usage of the Java class {\tt Canvas} in game programming; the second video introduce the polymorphism concept of Java; the third video shows the getter and setter functions in Java.

\vspace{0.1cm} \noindent \textbf{Questionnaire Design.}
To evaluate whether our prototype can help developers navigate and explore programming videos effectively, we designed a questionnaire for each video.
Each questionnaire has five questions (see Table~\ref{tbl:app2}).
The questions are designed based on our programming experiences and a survey of two developers, with the goal to cover different kinds of information including API usage, source code content, program output, and workflow that tutorial watchers may be interested in when learning a programming tutorial.

To answer these questions, participants in our user study have to watch, explore and summarize the source code written in the video and the process of writing the code.
In particular, participants are asked to summarize the opened and viewed Java class files for each video (V1/V2/V3-Q1).
In addition, they are also asked to identify some specific content in the videos, for example, the files that contain the main entry in the first video (V1-Q2), the subclass of the class Bank in the second video (V2-Q2), and the return value of different functions (V2-Q3/Q4/Q5).
As only the first video contains some complex APIs (i.e., APIs for GUI programming), we design two questions for the first video that ask participants to identify which APIs are used to set the size of {\tt JFrame} and {\tt Canvas} (V1-Q4/Q5).

\begin{figure} \vspace{-0.2cm}
\begin{lstlisting}
class Ocr{
    public int height;

    public void setHeight(int height){
        this.height = height;
    }

    publiic int getHeight(int x){
        this.height = x;
        return height;
    }
}
\end{lstlisting}
\caption{The initial version of getHeight and setHeight}\label{fig:code1}
\end{figure}

\begin{figure}\vspace{-0.2cm}
\begin{lstlisting}
class Ocr{
    public int height;

    public void setHeight(int height){
        if(height<10){
	        this.height=height;
	        System.out.println("Orc met criteria"); }
	    else{
            System.out.println(" Please enter a height under 10 feet");
        }
    }

    publiic int getHeight(){
        return height;
    }
}
\end{lstlisting}
\caption{The revised version of getHeight and setHeight}\label{fig:code2}
\end{figure}

A unique property of programming videos is to demonstrate the process of completing a programming task.
For example, at the beginning of the third video (V3), the tutorial author declares a function {\tt getHeight(int x)}, which assigns a value to the field {\tt height} and returns the value of {\tt height}.
Then the author adds a function {\tt setHeight(int height)} to set the value of {\tt height} and revises the {\tt getHeight()} by removing the parameter {\tt int x} and the assignment statement for the field {\tt height}.
Finally, the author makes {\tt setHeight(int)} set the value of {\tt height} with the input parameter only if the input parameter is less than 10, otherwise it sets the value of {\tt height} as 0.
Fig.~\ref{fig:code1} and Fig.~\ref{fig:code2} present the initial and the revised source code of the function {\tt getHeight} and {\tt setHeight}, respectively.}
For the third video, we designed four questions (V3-Q2/Q3/Q4/Q5) that require participants to properly understand the process (i.e., sequence of steps ) in a programming video in order to answer the questions correctly.

The first author developed standard answers to the questionnaires, which were further validated by the second and third author.
A small pilot study with three developers (one for each tutorial) was conducted to test the suitability and difficulty of the tutorials and questionnaires.
The complete questionnaires and their answers can be found in the website of our prototype tool\footnote{http://baolingfeng.xyz:5000}.

\vspace{0.1cm} \noindent \textbf{Participants.}
We recruited 10 undergraduate students from the College of Computer Science in Zhejiang University.
Out of these 10 students, 6 are junior students (freshman and sophomore) and 4 are senior students (junior and senior).
All 10 participants are not familiar with the Java programming tasks used in the user study.


We adopted between-subject design in our user study.
All the participants are required to complete the questionnaires for the three programming videos.
We divided 10 participants into two groups: the experimental group whose participants use the prototype of \emph{psc2code}-enhanced video player, and the control group whose participants use a regular video player. Each group has 3 junior participants and two senior participants.

\vspace{0.1cm} \noindent \textbf{Procedure.}
Before the user study, we gave a short tutorial on the features of \emph{psc2code} to participants in the experimental group. The training focuses only on system features.
We did not give the tutorial for regular video player as all the participants are familiar with how to use a video player.
We divided the whole user study into three sessions. In each session, the participants in the two groups are required to complete the questionnaire for one programming video tutorial.
At the beginning of an experiment session, a questionnaire web page is shown to the participants.
Once the participants click the start button, a web page is opened in another browser window or tab. The participants in the two groups use the corresponding tool to watch the video tutorial. When the participants complete the questionnaire, they submit the questionnaire by clicking the submit button.

After submitting the questionnaire, the participants in the two groups are asked to rate the usefulness of the corresponding tool (\emph{psc2code}-enhanced video player or the regular video player) they use for navigating and exploring the information in the programming video.
All the scores rated by participants are on a 5-points likert scale (1 being the worst, 5 being the best).
The participants can also write some suggestions or feedbacks in free text for the tool they use.
We mark the the correctness of the answers against the ground-truth answers, which is built when designing the questionnaires. The questionnaire website can calculate the completion time for each participant automatically.

\subsubsection{Results}
Table~\ref{tbl:userstudy1}, Table~\ref{tbl:userstudy2} and Table~\ref{tbl:userstudy3} present the results of the user study for the three programming videos respectively.
As shown in these tables, the average completion time of the participants using \emph{psc2code}-enhanced video player on the three videos are all less than that of the participants using the regular video player.
Furthermore, the average answer correctness of the participants using \emph{psc2code}-enhanced video player is all greater than that of the participants using the regular video player.
Since the programming tasks in the three programming videos and the questions about the information in the videos are not very complicated, the difference between the correctness of the participants using the two tools is not very large, except for the workflow related questions of the third video.
This suggests that it is difficult to navigate through the workflow and find the workflow-related information in the programming videos using the regular video player.
In contrast, \emph{psc2code}-enhanced video player can help video watchers navigate and explore the workflow information more efficiently and accurately.


All the participants in the experimental group agree that \emph{psc2code}-enhanced video player can help them navigate and explore the programming videos and learn the knowledge in the video tutorial effectively. One example of positive feedback is ``\textit{I can navigate the video by searching a specific code element, which help me find the answer easily.}'' Some participants also give some feedback on the drawbacks of our prototype tool, for example, ``\textit{some code is not complete (such as missing the bracket '\}' at the end of the line)}'' and ``\textit{there is a bit synchronization latency between the video playing and the update of the file content view}''.
But they also mention that these drawbacks have no significant impact on understanding the video tutorial.
In contrast, the ratings for the regular video player are not high. A participant complains that ``\textit{I have to watch the video very carefully because I do not want to miss some important information. Otherwise, I may have a headache to find what I miss and where it is in the video}''.

\begin{table}[]
    \caption{The average completion time, the answer correctness and the usefulness ratings by the five participants of the two groups for the video V1.}\label{tbl:userstudy1}
    \begin{tabular}{@{}ll|ll|ll@{}}
    \toprule
    \multicolumn{2}{c|}{Completion Time} & \multicolumn{2}{c|}{Correctness} & \multicolumn{2}{c}{Usefulness Rating}    \\ \midrule
    baseline               & psc2code              & baseline        & psc2code       & baseline     & psc2code     \\ \midrule
    353                    & 548                   & 1               & 0.8            & 2            & 4            \\
    86                     & 123                   & 1               & 1              & 1            & 5            \\
    404                    & 476                   & 0.8             & 1              & 3            & 5            \\
    583                    & 216                   & 1               & 1              & 4            & 4            \\
    409                    & 251                   & 0.8             & 1              & 2            & 5            \\ \midrule
    \textbf{367.0}         & \textbf{322.8}        & \textbf{0.92}   & \textbf{0.96}  & \textbf{2.4} & \textbf{4.6} \\ \bottomrule
    \end{tabular}
\end{table}

\begin{table}[]
    \caption{The average completion time, the answer correctness and the usefulness ratings by the five participants of the two groups for the video V2.}\label{tbl:userstudy2}
    \begin{tabular}{@{}ll|ll|ll@{}}
    \toprule
    \multicolumn{2}{c|}{Completion Time} & \multicolumn{2}{c|}{Correctness} & \multicolumn{2}{c}{Usefulness Rating}    \\ \midrule
    baseline               & psc2code              & baseline        & psc2code       & baseline     & psc2code     \\ \midrule
    408                    & 584                   & 1               & 1              & 3            & 4            \\
    212                    & 149                   & 1               & 0.8            & 1            & 5            \\
    1309                   & 318                   & 0.8             & 1              & 3            & 4            \\
    396                    & 313                   & 1               & 1              & 2            & 4            \\
    337                    & 140                   & 0.8             & 1              & 2            & 4            \\ \midrule
    \textbf{532.4}         & \textbf{300.8}        & \textbf{0.92}   & \textbf{0.96}  & \textbf{2.2} & \textbf{4.2} \\ \bottomrule
    \end{tabular}
\end{table}

\begin{table}[]
    \caption{The average completion time, the answer correctness and the usefulness ratings by the five participants of the two groups for the video V3.}\label{tbl:userstudy3}
    \begin{tabular}{@{}ll|ll|ll@{}}
    \toprule
    \multicolumn{2}{c|}{Completion Time} & \multicolumn{2}{c|}{Correctness} & \multicolumn{2}{c}{Usefulness Rating}    \\ \midrule
    baseline               & psc2code              & baseline       & psc2code        & baseline     & psc2code     \\ \midrule
    1083                   & 859                   & 0.8            & 0.8             & 2            & 4            \\
    379                    & 200                   & 0.8            & 1               & 3            & 5            \\
    913                    & 415                   & 1              & 1               & 2            & 5            \\
    709                    & 435                   & 0.8            & 0.8             & 3            & 3            \\
    489                    & 167                   & 0.6            & 1               & 1            & 4            \\ \midrule
    \textbf{714.6}         & \textbf{415.2}        & \textbf{0.8}   & \textbf{0.92}   & \textbf{2.2} & \textbf{4.2} \\ \bottomrule
    \end{tabular}
\end{table}


\subsubsection{Comparison with CodeMotion}


CodeMotion~\cite{khandwala2018codemotion} is a notable work, which is similar to our application for enhancing programming tutorial interaction. First, CodeMotion uses image processing techniques to identify potential code segments within frames. Then, it extracts text from segments using OCR and removes segments that don't look like source code. Finally, it detects code edits by computing differences between consecutive frames and splits a video into a set of 
time intervals based on chunks of related code edits, which allows users to view different phases in the video.
CodeMotion also has a web-based interface, in which a programming video are split into multiple mini-videos that is corresponding to the code edit intervals.

\begin{figure*}[t]
	\centering
	\includegraphics[width=0.7\textwidth]{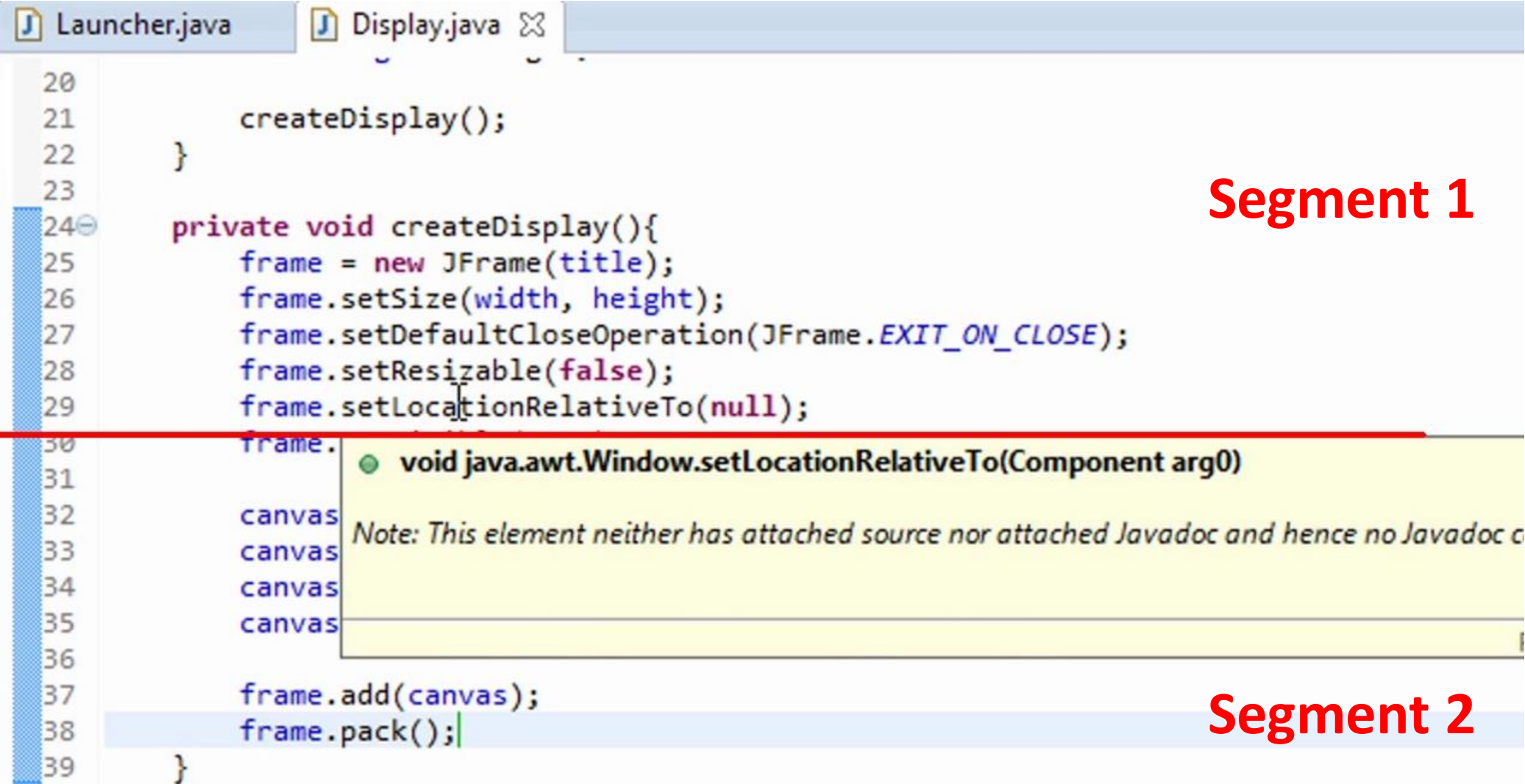}
	\caption{An example of segments of a frame generated by CodeMotion.}\label{fig:codemotion}
\end{figure*}

However, we think CodeMotion has several weaknesses, which are as follows: 
\begin{itemize}
		\item It doesn't reduce non-informative frames. CodeMotion extracts the first frame of each second and apply its segmentation algorithm and OCR on all extracted frames. 
		\item It doesn't remove noisy-code frames. CodeMotion filters segments that is not likely to have source code by identifying keywords of programming languages in the extracted text. But it can only remove the non-code segments. The noisy-code frames will affect the effectiveness of the segmentation algorithm and introduce errors in its follow-up steps. For example, Figure~\ref{fig:codemotion} presents an example of segments of a frame from V1 generated by CodeMotion. As shown in this figure, due to the popup windows, CodeMotion detects two segments and infers that these two segments contain source code. 
		\item It doesn't correct errors in the OCRed source code. CodeMotion only eliminates left-aligned text that looks
		like line numbers and converts accented characters or Unicode variants into their closest unaccented ASCII versions. However, many other OCR errors are not corrected. For example, among the OCR results of the video V2, there are many errors in the OCRed source code such as ``package'' $\rightarrow$ ``erackage'', ``int'' $\rightarrow$ ``bnt'', etc. 
\end{itemize}

We use the code of CodeMotion\footnote{Their code is host in GitHub (https://github.com/kandarpksk/codemotion-las2018), but is not well documented and maintained.} to process the three videos in our user study. Table~\ref{tbl:codemotion} shows the number of frames processed by CodeMotion and psc2Code. In this table, the column {\tt Total} is the number of frames processed by the segmentation algorithm and OCR technique of CodeMotion; the column {\tt Filtered} is the number of frames after CodeMotion removes segments that are not likely to have source code; the column {\tt \#valid} and {\tt \#invalid} are the number of valid and invalid frames generated by CodeMotion and psc2code, respectively. 
As shown, CodeMotion applied their segmentation algorithm and OCR technique on much more frames than our approach psc2code. Only a small number of frames that don't have source code are removed by CodeMotion. Among these remaining frames, there are many invalid frames.

The implementation of CodeMotion has a web-based interface, which presents a programming video as multiple mini-videos that is corresponding to the code edit intervals detected by it. Each mini-video corresponds a code edit interval, which is identified when: 1) the programming language of the detected code changes, or 2) the inter-frame
diff shows more than 70\% of the lines differing. Since only Java is used in the three videos used in the study, the code edit intervals would be identified when the inter-frame diff is large. However, the noisy-code frames have a big impact on the detection of the code edit intervals. For example, CodeMotion identified 8 code edit intervals for the video V1, but there were three code edit intervals that were caused by a popup window when running the program. 
Additionally, due to the errors in the OCRed source code, we find that it is difficult for users to use CodeMotion to navigate and understand the videos.
Comparing to CodeMotion, our tool psc2code detected three Java files from the video and showed them in a file content view, which is easier to navigate and understand. 
Thus, we believe that our tool have better user experience that CodeMotion.

\begin{table}[]
    \centering
    \caption{The number of frames processed by CodeMotion and psc2Code for the three videos in the user study.}\label{tbl:codemotion}
    \begin{tabular}{@{}l|rrrr|rr@{}}
    \toprule
       & \multicolumn{4}{c|}{CodeMotion}        & \multicolumn{2}{c}{psc2code} \\ \cmidrule(l){2-7} 
       & Total & Filtered & \#valid & \#invalid & \#valid      & \#invalid     \\ \midrule
    V1 & 353   & 321      & 252     & 69        & 54           & 20            \\
    V2 & 495   & 491      & 437     & 54        & 63           & 10            \\
    V3 & 673   & 672      & 645     & 27        & 65           & 16            \\ \bottomrule
    \end{tabular}
\end{table}

\section{Threats to Validity}\label{sec:diss}

\medskip \noindent \textbf{Threats to internal validity} refer to factors internal to our studies that could have influenced our experiment results:

\begin{enumerate}[leftmargin=*]
    \item \emph{The implementation and evaluation of psc2code:}
    We manually label and validate the non-code and noisy-code frames to build and evaluate the CNN-based image classifier used in \emph{psc2code}. There could be some frames that are erroneously labelled.
    To minimize human labeling errors, two annotators label and validate the frames independently, and their labels reach almost perfect agreement.
    In our current implementation, \emph{psc2code} extracts the largest detected rectangle as code editor sub-window based on our observation that most of valid code frames contain only a single code editor sub-window.
    However, we observe that developers in a video tutorial occasionally show two or more source code files side-by-side in the IDE.
    Our current simple largest-rectangle heuristic will miss some code regions in such cases.
    However, our approach can be easily extend to extract multiple code regions using a CNN-based image classifier to distinguish code-regions from non code-regions.

    To evaluate the quality of the \emph{psc2code}'s data processing steps, we randomly select two videos for each playlist. As developers usually use the same development environment to record video tutorials in a playlist, we believe that the analysis results for the two selected videos are representative for the corresponding playlist.
    When we have to manually examine the output quality of a data processing step, we always use two annotators and confirm the validity of data annotation by inter-rater agreement.
    In our experiments, the two annotators always have almost perfect agreement.

    \item \emph{The ground truth for locating code regions in code frames:}
    In RQ3, we use \emph{psc2code} to generated the ground truth to reduce the significant human effort and time required to annotate the ground-truth code region bounding boxes. However, a bias could potentially be introduced by the fact that the annotators (who are the first two authors) were aware of the prediction before defining the ground truth. The boundaries of code editor windows are easy to be recognized by annotators manually and we allow annotators to adjust the bounding box using a web interface. When the predicted code area is very close to the ground truth, annotators usually do not change the bounding box and the resulting IoU difference would actually be very small. When the difference between the predicted code area and the ground truth is big, annotators can adjust the predicted bounding boxes. In many cases, there is at least one border of the bounding boxes that matches the ground truth, which can help annotators match the ground truth. Thus, using predicted code area by \emph{psc2code} can save a lot of human effort and time.
    However, this annotation process may introduce some bias. To validate whether the difference between the ground truth and the results annotated by us is small and the introduced bias is acceptable, we randomly select 100 frames from our dataset and invite two graduate students to label the boundaries of code editor windows manually. Then, we compute the IoU between the ground truth generated by psc2code and the bounding box of the manual annotation for each frame. The average IoU is 0.96, which shows that the introduced bias is acceptable.

    \item \emph{Programming video search:} There might be some biases in the 20 queries that are used to evaluate the video search engine. We select these queries based on the APIs and programming concepts covered in our dataset of programming videos.
        More experiments are needed to generalize our results for different queries and video datasets.
        We manually check whether the videos in the returned list are truly relevant to the query, which might introduce human errors in the results. To minimize the errors, two annotators validate the results independently and their annotations reach almost perfect agreement.

    \item \emph{Enhancing programming video interaction:} 
    There might be some biases in the three selected videos for user study. To reduce the biases, we select three videos that contain different programming tasks from different playlists.
    The questionnaires for the three videos are designed collaboratively by the authors, with the goal to cover different categories of knowledge that developers could be interested in the programming tutorials.
    We also conduct a pilot study with two developers (different from the study participants) to make necessary revisions of questionnaires based on their feedbacks on the suitability and difficulty of tutorials and questionnaires.
    The participants cannot answer the questions without watching the videos, because all questions require participants to find relevant information in the video tutorials, rather than depending on their general programming knowledge.
    However, there might be some expectation biases that favor our prototype in the questionnaires.
    Another threat is that the limited number of participatns (i.e., 10) in the user study  might affect our conclusion. In the future, we plan to recruit more participants to investigate the effectiveness of the prototype. 

\end{enumerate}

\medskip \noindent \textbf{Threats to external validity} refer to the generalizability of the results in this study. We have applied \emph{psc2code} on 1042 Java video tutorials from YouTube, but those video tutorials do not cover all knowledge in Java. In the future, we plan to collect more video tutorials with different programming languages to further evaluate our \emph{psc2code} system and the downstream applications it enables.

\section{Related Work}\label{sec:related}
\subsection{Information Extraction in Screen-captured Videos}
Screen-captured techniques are widely used to record video tutorials. Researchers have proposed many approaches to extract different kinds of information from screen-captured videos (i.e., screencasts).

Some approaches (e.g., Prefab~\cite{dixon2010prefab}, Waken~\cite{banovic2012waken}, Sikuli~\cite{yeh2009sikuli}) use the computer vision technique to identify GUI elements in screen-captured images or videos. For example, Waken~\cite{banovic2012waken} use the image differencing technique to identify the occurrence of cursors, icons, menus, and tooltips that an application contains in a screencast. Sikuli~\cite{yeh2009sikuli} uses the template matching techniques to find GUI patterns on the screen. It also uses an OCR tool to extract text in the screenshots to facilitate video-content search.
The GUI elements identified by these approaches might be used to remove the noisy-code frames, for example, the completion suggestion popups usually have some UI patterns. But it is difficult to pre-define and recognize these UI patterns for the GUI windows with very diverse styles in different programming videos. Thus, in this study, we leverage a deep learning technique to remove the noisy-code frames.

Bao \emph{et al.}~\cite{bao2015reverse} proposed a tool named scvRipper to extract time-series developers' interaction data from programming screencasts. They model the GUI window based on the window layout and some special icons for each software application. Then, they identify an application using the image template techniques and crop the region of interest from the screenshots based on the model of the GUI windows.
Finally, they use an OCR tool to extract the textual content from the cropped images and construct developers' actions based on the OCRed textual content. However, it is impossible to model all the application windows in a large video dataset. In our study, we identify the code region based on the denoised horizontal and vertical lines that demarcate the boundaries of the main code editor window. 
Bao \emph{et al.}~\cite{bao2018vt} proposed another tool named VTRevolution to enhance programming video tutorials by recording and abstracting the workflow of developer when they create them. But their tool cannot be applied on existing programming video tutorials since it requires developers' interaction data recorded by a specific instrumentation tool~\cite{bao2015activityspace}.

Some tools such as NoteVideo~\cite{monserrat2013notevideo} and Visual Transcripts~\cite{shin2015visual} focus on specialized video tutorial, i.e., hand-sketched blackboard-style lecture videos popularized by Khan Academy~\cite{guo2014video}. They use computer vision techniques to identify the visual content in a video lecture as discrete visual entities including equations, figures, or lines of text. Different from these tools, our \emph{psc2code} focuses on the programming video tutorials where developers are writing code in IDEs.

\subsection{Source Code Detection and Extraction in Programming Screencasts}
One notable approach to extract source code from programming videos is CodeTube~\cite{ponzanelli2016too} by Ponzanelli et al., which is a web-based recommendation system for programming video tutorial search based on the extracted source code.
To remove the noise for code extraction, CodeTube uses shape detection to identify code regions, it does not attempt to reduce the noisy edge detection results as our approach does.
It identifies Java code by applying OCR to the cropped code-region image followed by using an island parser to extract code constructs.
CodeTube also identifies video fragments characterized by the presence of a specific piece of code.
If the code constructs in a video fragment are not found to be similar, it applies a Longest Common Substring (LCS) analysis on image pixels to find frames with similar content.
Also, Ponzanelli et al. divided video fragments into differnt categories(e.g., theoretical, implementation) and complements the video fragments with relevant Stack Overflow discussions~\cite{ponzanelli2017automatic}.
Different from the CodeTube's post-processing of the OCRed source code, our approach clusters frames with the same window layout to denoise the noisy candidate window boundary lines before identifying and cropping code regions in frames.
Furthermore, our approach does not simply discard inconsistent OCRed source code across different frames, but tries to fix the OCR errors with cross-frame information of the same code.


Similar to our work, Ott et al.~\cite{OttAHBL08} proposed to use a VGG network to identify whether frames in programming tutorial videos contain source code. They also use deep learning techniques to classify images based on programming language~\cite{ott2018learning} and UML diagrams~\cite{ott2019exploring}. 
In our study, we combine deep learning techniques and traditional computer vision techniques to achieve better performance than Ott et al.'s approach.
Yadid and Yahav~\cite{yadid2016extracting} also wanted to address the issue that the errors in OCRed source code result in a low precision in searching code snippet in programming videos. They used cross-frame information and statistical language models to make corrections by selecting the most likely token and line of code. They conducted an experiment on 40 video tutorials, and found that their approach can extract source code from programming videos with high accuracy. Khandwala and Guo~\cite{khandwala2018codemotion} also used the computer vision technique to identify code from programming videos. But their focus was to use the extracted source code to enhance the design space of interactions with programming videos.
Moslehi et al. used the extracted text from screencasts to perform feature location tasks~\cite{moslehi2018feature}.

In our study, we not only follow the approach proposed by Yadid and Yahav~\cite{yadid2016extracting} to make corrections in the OCRed source code, but also leverage a deep learning technique to remove non-code and noisy-code frames before OCRing the source code from frames. This is because we find that it is difficult to remove the noise in some kinds of frames (such as the frames with completion suggestion popups) using traditional computer vision techniques as in~\cite{bao2015reverse} or the post-processing of the OCRed code as in~\cite{ponzanelli2016too}.
Inspired by the work of Ott et al.~\cite{OttAHBL08}, we develop a CNN-based image classifierto identify non-code and noisy-code frames.
Ott et al.~\cite{OttAHBL08} train a deep learning model to identify the presence of source code in thousands of frames. The deep learning model in their study can identify four categories of frames: \textit{Visible Typeset Code, Partially Visible Typeset Code, Handwritten Code, and No Code}, and achieves very high accuracies (85.6\%-98.6\%).
We follow their approach but the number of classes in our task is 2 (i.e., valid and invalid) and train a model using our own dataset.

\section{Conclusion}\label{sec:conclusion}
In this paper, we develop an approach and a system named \emph{psc2code} to denoise source code extracted from programming screencasts.
First, we train a CNN-based image classifier to predict whether a frame is a valid code frame or non-code or noisy-code frame. After removing non-code/noisy-code frames, \emph{psc2code} extracts the code regions based on the detection of sub-window boundaries and the clustering of frames with the same window-layout.
Finally, \emph{psc2code} uses a professional OCR tool to extract source code from videos and leverage the cross-frame information in a programming screencast and the statistical language model of a large source-code corpus to correct the OCR errors in the OCRed source code.

We collect 23 playlists with 1,142 programming videos from YouTube to build a programming-video dataset used in our experiments. 
We systematically evaluate the effectiveness of the four main steps of \emph{psc2code} on this video dataset. Our experiment results confirm that the denoising steps of \emph{psc2code} can significantly improve the quality of source code extracted from programming screencasts.
Based on the denoised source code extracted by \emph{psc2code}, we implement two applications.
First, we build a programming video search engine. We use 20 queries of some commonly used Java APIs and programming concepts to evaluate the video search engine on the denoised source-code corpus extracted by \emph{psc2code} versus the noisy source-code corpus extracted without using \emph{psc2code}. 
The experiment shows that the denoisied source-code corpus enables a much better video search accuracy, compared with the noisy source-code corpus.
Second, we build a web-based prototype tool to enhance the navigation and exploration of programming videos based on the \emph{psc2code}-extracted source code. We conduct a user study with 10 participants and find that the \emph{psc2code}-enhanced video player can help participants navigate the programming videos and find content-, API-usage- and process-related information in the video tutorial more efficiently and more accurately, compared with using a regular video player.

\begin{acks}
  This research was partially supported by the National Key Research and Development Program of China (2018YFB1003904), NSFC Program (No. 61972339), NSFC Program (No. 61902344), the Australian Research Council's Discovery Early Career Researcher Award (DECRA) funding scheme (DE200100021), and ANU-Data61 Collaborative Researh Project CO19314.
\end{acks}

\bibliographystyle{ACM-Reference-Format}
\bibliography{reference}

\end{document}